\documentclass[sigconf, authorversion]{acmart}

\usepackage{multirow}
\usepackage{acronym}
\usepackage{graphicx}
\usepackage{subcaption}
\usepackage{comment}
\usepackage{balance}
\usepackage{amsthm}
\usepackage{tabularx}
\usepackage{color}
\usepackage{hyperref}
\usepackage{pifont}
\usepackage{adjustbox}

\newcommand{\name}{\emph{BrokenStrokes}}
\newcolumntype{P}[1]{>{\centering\arraybackslash}m{#1}}
\newcommand{\cmark}{\ding{51}}
\newcommand{\xmark}{\ding{55}}
\AtBeginDocument{%
  \providecommand\BibTeX{{%
    \normalfont B\kern-0.5em{\scshape i\kern-0.25em b}\kern-0.8em\TeX}}}

\acmConference[WISEC '20]{WISEC '20: 13th ACM Conference on Security and Privacy in Wireless and Mobile Networks}{Jul. 8--10, 2020}{Linz, AT}

\acrodef{RSS}{Received Signal Strength}
\acrodef{SDR}{Software Defined Radio}
\acrodef{LOS}{Line-Of-Sight}
\acrodef{NLOS}{Non-Line-of-Sight}
\acrodef{RF}{Radio Frequency}
\acrodef{SVM}{Support Vector Machine}
\acrodef{RF}{Radio Frequency}
\acrodef{TP}{True Positive}
\acrodef{FP}{False Positive}
\acrodef{FN}{False Negative}
\acrodef{SSH}{Secure Shell}
\acrodef{ML}{Machine Learning}
\acrodef{CSI}{Channel State Information}
\acrodef{FHSS}{Frequency Hopping Spread Spectrum}
\acrodef{DSSS}{Direct Sequence Spread Spectrum}

\begin{document}

%
\title{BrokenStrokes: On the (in)Security of Wireless Keyboards}

\author{Gabriele Oligeri, Savio Sciancalepore, Simone Raponi, Roberto Di Pietro}
\affiliation{
  \institution{Division of Information and Computing Technology \\ College of Science and Engineering, Hamad Bin Khalifa University}
  \state{Doha, Qatar}
}

\acmYear{2020}\copyrightyear{2020}
\setcopyright{acmcopyright}
\acmConference[WiSec '20]{13th ACM Conference on Security and Privacy in Wireless and Mobile Networks}{July 8--10, 2020}{Linz (Virtual Event), Austria}
\acmBooktitle{13th ACM Conference on Security and Privacy in Wireless and Mobile Networks (WiSec '20), July 8--10, 2020, Linz (Virtual Event), Austria}
\acmPrice{15.00}
\acmDOI{10.1145/3395351.3399351}
\acmISBN{978-1-4503-8006-5/20/07}

%
\begin{abstract}
Wireless devices resorting to event-triggered communications have been proved to suffer critical privacy issues, due to the intrinsic leakage associated with radio-frequency (RF) emissions.

In this paper, we move the attack frontier forward by proposing \name: an inexpensive, easy to implement, efficient, and effective attack able to detect the typing of a pre-defined keyword by only eavesdropping the communication channel used by the wireless keyboard. \name\ proves itself to be a particularly dreadful attack: it achieves its goal when the eavesdropping antenna is up to 15 meters from the target keyboard, 
regardless of the encryption scheme, the communication protocol, the presence of radio noise, and the presence of physical obstacles. 
While we detail the attack in three current scenarios and discuss its striking performance---its success probability exceeds  $90\%$ in normal operating conditions---, we also provide some suggestions on how to mitigate it.
The data utilized in this paper have been released as open-source to allow practitioners, industries, and academia to verify our claims and use them as a basis for further developments.

\end{abstract}




\maketitle

\section{Introduction}
\label{sec:introduction}

Wireless keyboards are becoming more and more popular in homes, offices, and entertainment systems, enabling a smooth, tiny, and elegant interaction with computing devices~\cite{Ali2017_JSAC}. Especially in crowded offices, wireless keyboards reduce the number of wires to be managed per working location, with evident advantages in elegance and neatness. Besides, they extend the interaction area with terminals, allowing stress-less and pain-free working experiences~\cite{Boitan2018}. 


Despite their popularity, wireless keyboards suffer several confidentiality and privacy issues, mainly caused by the broadcast nature of the wireless communication link and energy constraints~\cite{Boitan2018}. In fact, compared to legacy wired keyboards, wireless keyboards use a wireless communication medium, where the information is inherently exposed to potential eavesdropping~\cite{Albazrqaoe2019}. At the same time, being powered by batteries, wireless keyboards have to implement efficient computation and communication strategies, minimizing the \ac{RF} operations to increase the lifetime of the batteries~\cite{Rault2014}.
From the security perspective, many legacy wireless keyboards deploy very weak (or none) protection against eavesdropping attacks. In the cited context, attacks can be easily achieved by tuning a malicious receiver at the same operating frequency of the keyboard~\cite{Chen2015_mobisys}.
A few researchers~\cite{keystrokeinjection} also demonstrated the feasibility of active attacks, such as keystroke injection and replay, capable of poisoning the communication link and reducing the usability and security of wireless keyboards. 
While manufacturers are designing, implementing and delivering more and more secure solutions for wireless keyboards, the intrinsic security of wireless keyboards has still to deal with usability and energy constraints~\cite{Halevi2015}. Indeed, wireless keyboards have to trigger a new \ac{RF} communication for each new keystroke, to guarantee the minimum typing delay and maximum usability. At the same time, such \ac{RF} communication should last for the minimum amount of time, to minimize the battery drain and increase the lifetime of the keyboard battery itself~\cite{Chen2015_mobisys}.

{\bf Contribution.} In this paper, we present \name, a novel attack able to detect the presence of specific  keywords in arbitrarily long keystroke sequences by only eavesdropping the (encrypted) keyboard-dongle communication link. The underlying strategy of \name\ is based on the identification and acquisition of \ac{RSS} samples associated with the keystrokes of a target user. 
Applying ML techniques to the eavesdropped encrypted traffic between the keyboard and the dongle, \name\ can enable a variety of attacks, including the identification of the number of keystrokes associated with a keyword, as well as the detection of a specific keyword in a stream of keystrokes. Overall, \name\ is a very inexpensive and easy-to-perform attack, requiring only a commercial \ac{SDR} and an antenna working on the 2.4 GHz frequency band. Moreover, \name\ is a completely agnostic attack, being independent of (i) MAC-layer communication protocol, (ii) packet format, and, (iii) the adopted encryption layer.


We stress that \name\ significantly improves keystroke eavesdropping and analysis compared to the current state of the art. Indeed, contrary to other related work (see Section~\ref{sec:relatetd_work} for an overview), \name\ is effective even when no information on the MAC-layer protocol is available, in the presence of obstacles, and up to distances of about 15 meters from the target keyboard. Besides, despite we show an application of \name\ in the area of keyword identification, we remark that the main novel contribution of our work is in the accurate translation between RSS recordings and keystrokes inter-arrival times. Then, \name\ can be easily coupled with any of the well-known keystroke analysis techniques available in the literature, to provide the desired objective (password guessing, keyword identification, text analysis, and so on).

{\bf Paper Organization.} The remainder of this paper is organized as follows: Section~\ref{sec:relatetd_work} summarizes recent attacks against keyboards, while Section~\ref{sec:scenario} illustrates our assumptions and the considered scenario. The intuition behind \name\ is introduced in Section~\ref{sec:attack_in_brief}, while Section~\ref{sec:time_extraction} describes the methodology to detect a keyword in a sequence of keystrokes, by exploiting the \ac{RSS}. The three scenarios tackled by our contribution are described in Sections~\ref{sec:scenario1},~\ref{sec:scenario2}, and~\ref{sec:scenario3}, respectively. Section~\ref{sec:performance} provides the results of our attack in all the cited scenarios, while Section~\ref{sec:discussion} provides further details on the feasibility of \name, as well as some limitations. Finally, Section~\ref{sec:conclusion} tightens conclusions and draws some future work.

\section{Related Work}
\label{sec:relatetd_work}

Keylogging side-channel attacks can be classified as a function of different parameters~\cite{monaco}, including \emph{targets} (user, keyboard, host or network), \emph{modality} (acoustic, wired, WiFi, seismic, motion, EM radiations), and \emph{proximity} (close proximity
, few meters, or up to 15 meters). In the following, we provide a brief overview of communication attacks---being \name\ in the same category.

This class of attacks explores the possibility of reconstructing the keystrokes typed by target users by extracting information from the communication channel, be it wired or wireless. Focusing on the wired setting, an early analysis of keystrokes timing attacks has been provided by~\cite{Song2001}. The authors collected inter-keystroke timings from Ethernet sessions using the \ac{SSH} protocol, and inferred on the bigrams typed by the user. The proposed solution allows to significantly reduce the entropy of passwords transmitted as encrypted via an SSH tunnel. While being characterized by outstanding performance, this solution requires physical access to the Ethernet link. Besides, it is suitable only for reducing the complexity of single word instances, such as passwords.
In the context of wireless communication attacks, the authors in~\cite{vuagnoux2009compromising} described the limitations of detecting compromised electromagnetic waves with a wide-band receiver tuned on a specific frequency. As a result, they proposed a new effective attack, consisting of acquiring the raw signal from the antenna and processing the entire electromagnetic spectrum. Despite being quite an expensive solution, this side-channel attack can recover 95\% of keystrokes on a PS/2 keyboard, from up to 20 meters, and through walls.
Similarly, the authors in~\cite{Ali2015_MCN} introduced a novel attack exploiting WiFi signals, which correlates the hand movement with text writing. When a user types a certain key, her fingers move uniquely, thus generating a unique pattern in the \ac{CSI}. The authors exploited WiFi signals to perform keystroke recognition by using two commercial devices: a sender (i.e., a router) and a receiver (i.e., a laptop). When evaluated in real-world experiments, the approach recognizes keystrokes with an accuracy of 93.5\%. A similar attack has been described by the authors in~\cite{Chen2015_mobisys}, based on the identification of the changes in the wireless channels related to a keystroke. By relying on five antennas and signal-cancellation techniques, the proposed solution reaches 91.8\% accuracy with full-training and 80\% accuracy with reduced training input. Eavesdropping attacks based on the \ac{CSI} extracted from wireless signals have emerged as effective strategies and can be delivered without relying on a training phase~\cite{fang2018no}.



\begin{table}[htbp]
    \footnotesize
	\caption{Comparing \name\ with related work.}
	\label{tab:comparison}
	\centering
	\begin{tabular}{|P{1.4cm}| P{1.6cm}| P{0.9cm} | P{1.6cm} | P{1cm}|}
		\hline
		\textbf{Ref.} & \textbf{MAC Protocol Agnostic} & \textbf{Long Range} & \textbf{Robustness to Obstacles} & \textbf{Reduced Cost} \\
		\hline
		\cite{Song2001} & \cmark & \xmark & \xmark & \cmark \\
		\hline
		\cite{vuagnoux2009compromising} & \cmark & \xmark & \xmark & \xmark \\
		\hline
		\cite{Ali2015_MCN} & \xmark & \cmark & \cmark & \cmark \\
		\hline
		\cite{Chen2015_mobisys} & \xmark & \cmark & \cmark & \cmark \\
		\hline
		\cite{fang2018no} & \xmark & \cmark & \cmark & \cmark \\
		\hline
		Our approach & \cmark & \cmark & \cmark & \cmark \\
		\hline
	\end{tabular}
\end{table}

As summarized in Table~\ref{tab:comparison}, compared to the above valuable approaches, \name\ is as a very flexible attack, agnostic respect to the communication protocol, being effective from 20cm up to 15m, and requiring minimal, cost-effective equipment.

\section{Scenario and assumptions}
\label{sec:scenario}

We consider a general scenario constituted by a wireless keyboard system, i.e., a keyboard transmitting wirelessly the user's keystrokes to a USB dongle connected to a computer. In this scenario, our attack affects all the wireless communication protocols that could be employed to sustain the communication between the keyboard and the dongle, such as Bluetooth, WiFi, and proprietary protocols. 
Without loss of generality, we consider three widely adopted wireless keyboards, as depicted in Table~\ref{tab:keyboards}. All of the keyboards feature proprietary communication protocols exploiting the ISM bandwidth $\left[ 2.4 - 2.5 \right]$ GHz for the communication. We stress that our solution involves neither the hacking nor the reverse engineering of the protocols adopted by the considered wireless keyboards. Moreover, we highlight that all the considered keyboards' brands implement encryption schemes that prevent direct access to the content of the exchanged messages.

\begin{table}[htbp]
    \footnotesize
	\caption{Considered keyboards---Brands and Models.}
	\label{tab:keyboards}
	\centering
    	\begin{adjustbox}{max width=0.95\columnwidth}
    	\begin{tabular}{|P{1cm}| P{1cm}| P{1.5cm} | P{2.5cm} |}
    		\hline
    		\textbf{Brand} & \textbf{Model} & \textbf{Frequency range [GHz]} & \textbf{Protocol / Security} \\
    		\hline
    		HP & SK-2064 & $\left[ 2.4 - 2.5 \right]$ & Proprietary / Encrypted\\
    		\hline
    		Microsoft & 850-1455 & $\left[ 2.4 - 2.5 \right]$ & Proprietary / Encrypted\\
    		\hline
    		V-MAX & K-201 & $\left[ 2.4 - 2.5 \right]$ & Proprietary / Encrypted\\
    		\hline
    	\end{tabular}
	\end{adjustbox}
\end{table}

{\bf Equipment.} We adopted a commercial laptop (Dell XPS 15 9560), featuring a Linux distribution and GNU Radio (a free and open-source software development toolkit), a commercial \ac{SDR}~\cite{Baldini2012}, and either an omnidirectional (VERT2450) or a directional antenna (Aaronia HyperLOG 60350), depending on the considered attack scenarios. Finally, all the proposed algorithms, techniques, and procedures adopted throughout this paper have been implemented in Matlab R2019a.


{\bf Scenario.} We performed extensive measurements of the \name\ attack in the following reference scenarios:
\begin{enumerate}
    \item {\bf Scenario 1: Proximity attack.} The \ac{SDR} features a standard omnidirectional antenna (VERT2450). We placed the \ac{SDR} in the close proximity of the keyboard-dongle communication link---we concealed it under the desk. This attack involves the adversary having access to the location of the target user (e.g., office, home), and being able to place the \ac{SDR} very close to the wireless keyboard, e.g., under the user's desk or in its close proximity.
    \item {\bf Scenario 2: Behind-the-wall attack.} The \ac{SDR} is connected to a directional antenna (Aaronia HyperLOG 60350) and there is no \ac{LOS} between the antenna and the keyboard-dongle communication link. This attack considers an adversary willing to collect the inter-keystrokes timings of a target user while being behind obstructing objects, such as walls~\cite{adib2015capturing, adib2013see}, thus possibly remaining undetected.
    \item {\bf Scenario 3: Remote attack.} In our setting the \ac{SDR} is connected to a long-range directional antenna (Aaronia HyperLOG 60350), the adversary is located far away from the target user, but has a clear \ac{LOS} to the target and can collect the inter-keystroke timings from a remote location (up to 15m).
\end{enumerate}

{\bf Multiple Users.} We considered three different users, namely $\{U1, U2, U3\}$, and we evaluated how the user's typing pace affects the \name\ attack. Note that the number of users considered in this paper is consistent with related works on keystrokes analysis~\cite{zhuang2009keyboard, zhu2014context, marquardt2011sp}.

{\bf Noise.} We remark that our measurement campaign has been performed in regular office conditions, without any effort to reduce the noise generated by other devices sharing the same communication frequency of the target keyboards.

{\bf Keyword dataset.} \name\ involves a two-stage attack, i.e., converting the received signal strength peaks to inter-keystroke timings, and then to keywords. While the vast majority of the literature focused on translating inter-keystroke timings to keywords by exploiting different physical layer hacks, we mainly focus on designing reliable and effective solutions to translate the received signal strength to timings. Without loss of generality, in this paper, we consider only one keyword, i.e., \emph{password}, being the second part of the attack an important, but not strictly novel contribution, to the current state of the art.

\section{\name\ in a nutshell}
\label{sec:attack_in_brief}


The computing flow of \name\ is composed of: (i) measuring the \acf{RSS} of the messages transmitted between the keyboard and the dongle; (ii) exploiting such measurements to extract inter-keystroke timings; and, finally, (iii) resorting to a \acf{ML} technique to generate a likelihood score, indicating the presence of a pre-defined keyword in the keystroke sequence of the target user. 

We adopted the MiriadRF LimeSDR to measure the \ac{RSS} of the packets generated by each keystroke event~\cite{Akeela2018}, and we resort to GNU Radio to tune the parameters of the SDR~\cite{Blossom2004}. Specifically, we observed that wireless keyboards are idle when no keystrokes are typed. As soon as the user presses any button, a new transmission from the keyboard to the dongle is triggered, generating a peak at a specific operating frequency.



We connected the \emph{LimeSDR Source (RX)} standard module, configured with a proper operating frequency, a bandwidth of 10 MHz, and a sample rate of 30 MHz, to a \emph{QT GUI Frequency Plot} module, where we enabled the log of the \ac{RSS} (in dBm) and a timestamp (in nanoseconds), when the value of the \ac{RSS} on the particular operating frequency exceeds a predefined threshold value. 




The log file generated by the \emph{Acquisition} module, containing the \ac{RSS} and the timestamps, is subsequently processed by a chain of Matlab scripts, i.e., \emph{Keystroke Timing Extraction} and a \ac{ML} algorithm, to generate the likelihood score associated with the presence of the keyword. The \emph{Keystroke Timing Extraction} block aims at identifying the keystroke patterns and generating the inter-keystroke timings, i.e., the time occurring between subsequent keystrokes of the user. Then, the interarrival times are passed to a \ac{ML} algorithm, which provides a likelihood score about the presence of a pre-defined keyword---the \ac{ML} algorithm has been previously used for training a model with different repetitions of the same keyword to be detected. More details of each phase involved in the \name\ attack will be provided in the next sections.

\section{From RSS to Keyword detection}
\label{sec:time_extraction}

In this section, we show the details of \name, providing the mechanisms that can be used by an adversary to detect the presence of a keyword in an arbitrarily long sentence typed by a user through a wireless keyboard. Without loss of generality, we consider Scenario 1, i.e., the \emph{Proximity attack}, while in the later sections, we will extend our methodology to the other scenarios.

\begin{figure}[htbp]
    \centering
    \includegraphics[width=0.8\columnwidth]{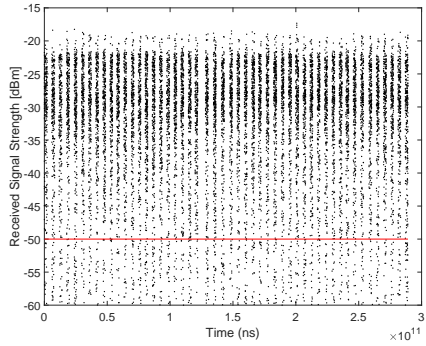}
    \caption{\acf{RSS} associated with 50 repetitions of the word ``password'' by user $U1$, assuming Scenario 1 (\emph{Proximity attack}).}
    \label{fig:samples_savio_50}
\end{figure}

Figure~\ref{fig:samples_savio_50} shows the \ac{RSS} samples collected from the \ac{SDR} with a sampling rate of $30^6$ samples per second. We asked $U1$ to type 50 times the keyword ``password'', and we collected the \ac{RSS} estimations associated with the messages exchanged between the keyboard and the dongle. We stress that the keyword ``password'' is not related to any specific user password. Indeed, it represents a generic 8-letters keyword that, without loss of generality, can be re-conducted to any keyword typed by the user in an arbitrarily long keystroke sequence, as detailed in the remainder of this paper.

The \ac{RSS} samples show a clear pattern, consisting of vertical bands: one band per word, since $U1$ was typing a keyword, hitting return, and then starting again---for a total of 50 repetitions of the word ``password''. The solid red line in Fig.~\ref{fig:samples_savio_50} shows the threshold we used for filtering \ac{RSS} samples, i.e., only the samples above the threshold are considered for the subsequent processing. The importance of the threshold will be clear later on, when filtering out the samples associated with interference while retaining only the samples coming from the keyboard-dongle communication. Indeed, we observe that, in this specific scenario (\emph{Proximity attack}), the vast majority of the samples are mainly concentrated in the range of $[-20, -35]$ dBm and, therefore, any threshold less than -35dBm can be adopted for this purpose.

In the following, we extract the inter-keystroke timings via a dual-stage process: (i) Words Identification; and, (ii) Keystroke Timings Extraction. The first phase exploits \ac{RSS} samples to identify the words typed by the target user, while the second phase focuses on extracting the inter-keystroke timings associated with the previously identified word.

\subsection{Words identification}
\label{sec:identification_of_words}

\begin{figure}[htbp]
    \centering
     \includegraphics[width=0.8\columnwidth]{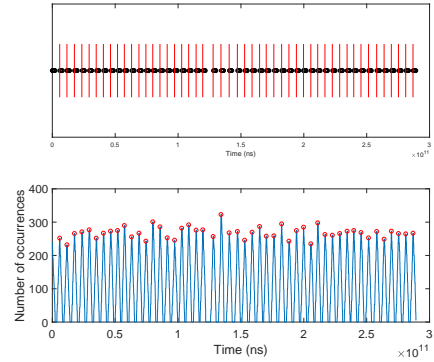}
    \caption{Words identification: we count for the number of RSS samples belonging to a sliding window (bottom figure) and we consider the beginning of the word at the peaks (top figure).}
    \label{fig:word_detection_savio}
\end{figure}

The top part of Fig.~\ref{fig:word_detection_savio} shows the samples collected for the experiment of Fig.~\ref{fig:samples_savio_50}, where all the RSS values have been normalized to the same value, being \ac{RSS} values not relevant for the subsequent analysis. To extract the timings associated with the beginning of each word, we considered a sliding window of a pre-determined duration, and we count for the number of samples belonging to it (when sliding from the beginning to the end of the trace). The sliding-window size is important and its configuration depends on both the user and the word to be detected. As an example, for the word ``password'', we considered sliding windows of size $2.4$, $1.7$, and $2$ seconds, for the user $U1$, $U2$, and $U3$, respectively. Moreover, we empirically assumed a sliding step of 1/50 of the window size. Finally, in this work, we assume that the sliding-window duration can be pre-set by the adversary. Indeed, it can properly calibrate the sliding window by looking at the collected samples, and set it up accordingly. The bottom part of Fig.~\ref{fig:word_detection_savio} shows the number of samples belonging to the sliding window, given a certain delay (in milliseconds) from the beginning of the trace. At the same time, the peaks in the bottom part of Fig.~\ref{fig:word_detection_savio} represent the beginning of a new word. Indeed, if the sliding window duration is properly calibrated, the number of samples is the highest possible when the window is at the beginning of the word. Vertical red lines in the top part of Fig.~\ref{fig:word_detection_savio} show the identified peaks in relation to the \ac{RSS} sample positions (black circles).

\subsection{Keystroke timings extraction}
\label{sec:keystroke_timing_extraction}

\begin{figure}[htbp]
    \centering
    \includegraphics[width=0.8\columnwidth]{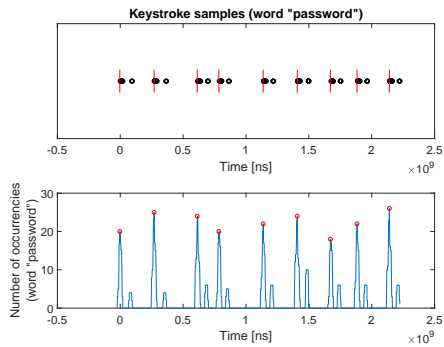}
    \caption{Keystroke timings extraction: we count for the number of RSS samples belonging to the sliding window (bottom figure) and we consider the peaks as the timings at which the keystrokes happen (vertical red lines in the top figure).}
    \label{fig:keystroke_sample_savio_word1}
\end{figure}

For each of the identified words (vertical red lines in Fig.~\ref{fig:word_detection_savio}), we performed the following analysis. Firstly, we focused on the samples collected from a single word, as depicted in the top part of Fig.~\ref{fig:keystroke_sample_savio_word1}. We observe that the word ``password'' is constituted by 9 groups of samples (8 letters and the carriage return). Each group of samples, in turn, can be divided into two sub-groups: the first set of about 20 samples and the second set of about 5 samples, as depicted in the bottom part of Fig.~\ref{fig:keystroke_sample_savio_word1}. We assume that the first sub-group belongs to the information packet transmitted by the keyboard to the dongle, while the second sub-group belongs to the acknowledgment message transmitted by the dongle to the keyboard. Our intuition is that each keystroke corresponds to one transmission by the keyboard and the corresponding acknowledgment message by the dongle. Without loss of generality, in the remainder of this section, we do not consider any packet loss between the keyboard and the dongle, consistently with the Scenario 1, where the eavesdropping equipment is very close to the keyboard-dongle communication link. Interference will be taken into account in later sections of this work (for scenarios 2 and 3), as well as strategies to mitigate their effect.
To correctly identify the keystroke timings, we adopted a sliding window duration of $0.024$ seconds and a sliding step of 1/50 of the window size. The sliding window duration takes into account the communication round-trip-delay between the keyboard and the dongle and, being dependent on the keyboard brand/model, it requires a pre-processing of the collected samples. The above parameters have been optimized for the HP SK-2064, while we will discuss the impact of the keyboard hardware on the performance of the \name\ attack in a later section of this paper (Sec. \ref{sec:discussion}).
Finally, by considering the peaks from the previous analysis, we identified the keystroke timings as depicted by the vertical red lines in the top part of Fig.~\ref{fig:keystroke_sample_savio_word1}. 

\begin{figure*}
    \begin{subfigure}{0.33\textwidth}
        \includegraphics[width=\linewidth,height=60mm]{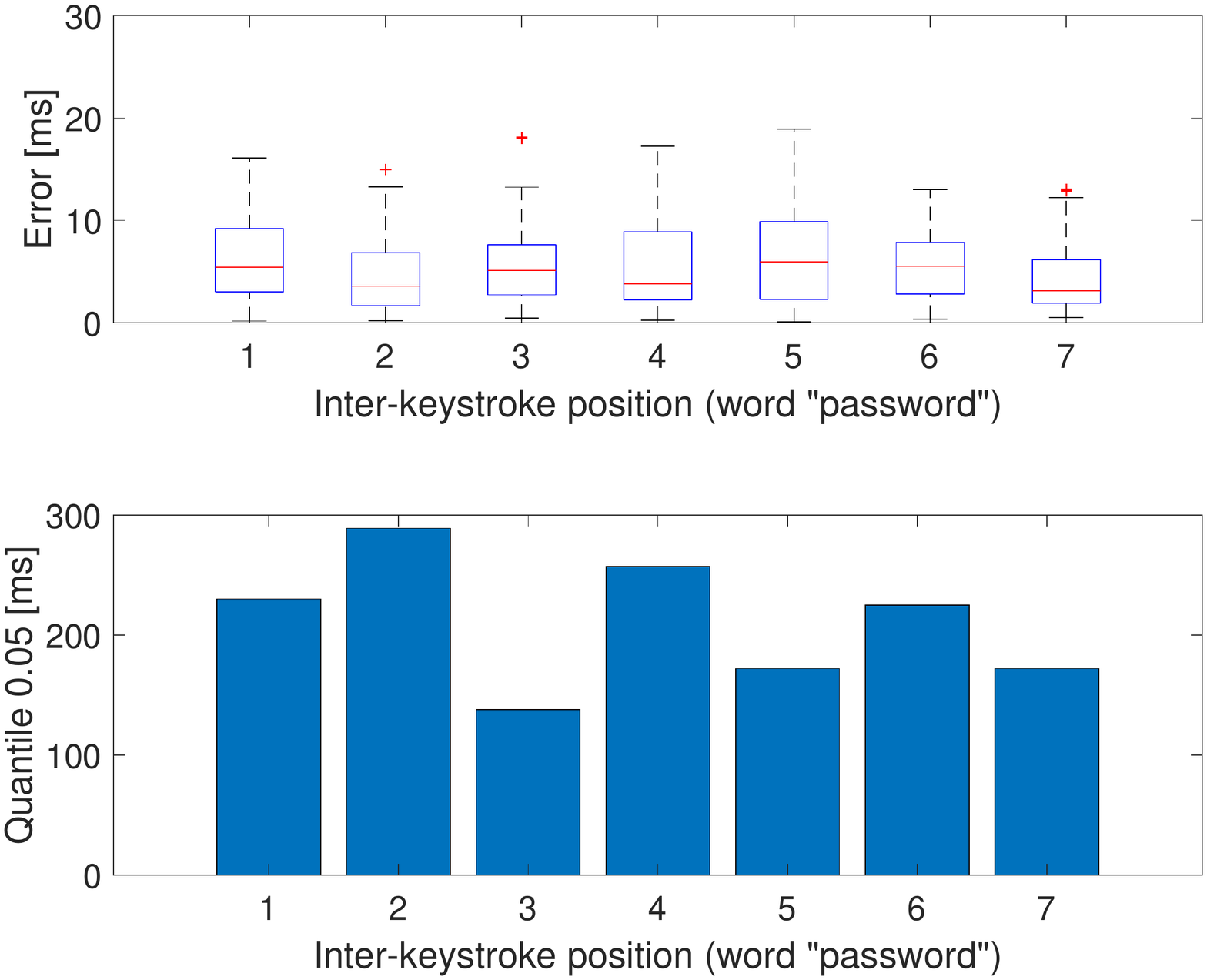}
        \caption{User 1}
    \end{subfigure}\hfill
    \begin{subfigure}{0.33\textwidth}
        \includegraphics[width=\linewidth,height=60mm]{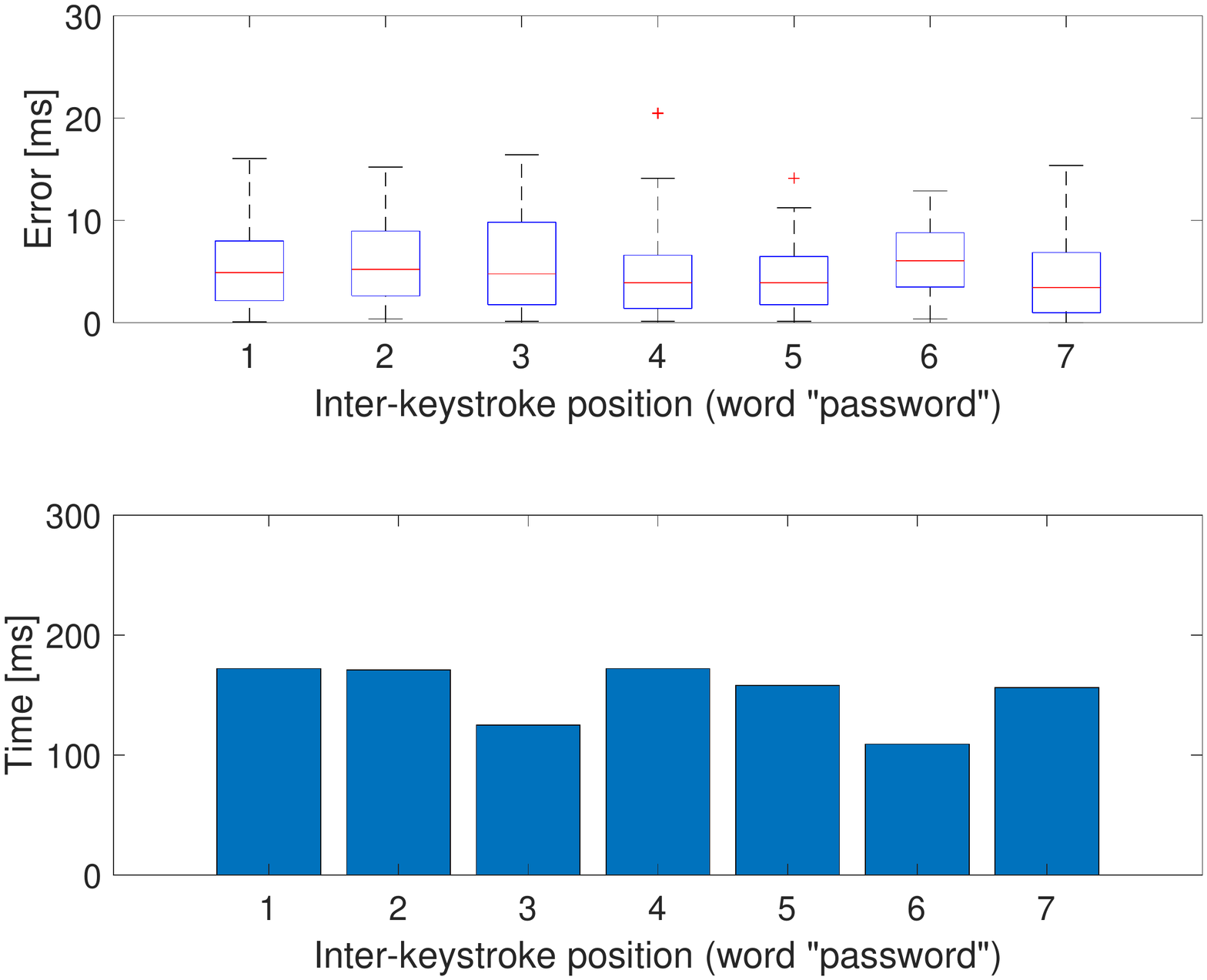}
        \caption{User 2}
    \end{subfigure}\hfill
    \begin{subfigure}{0.33\textwidth}%
        \includegraphics[width=\linewidth,height=60mm]{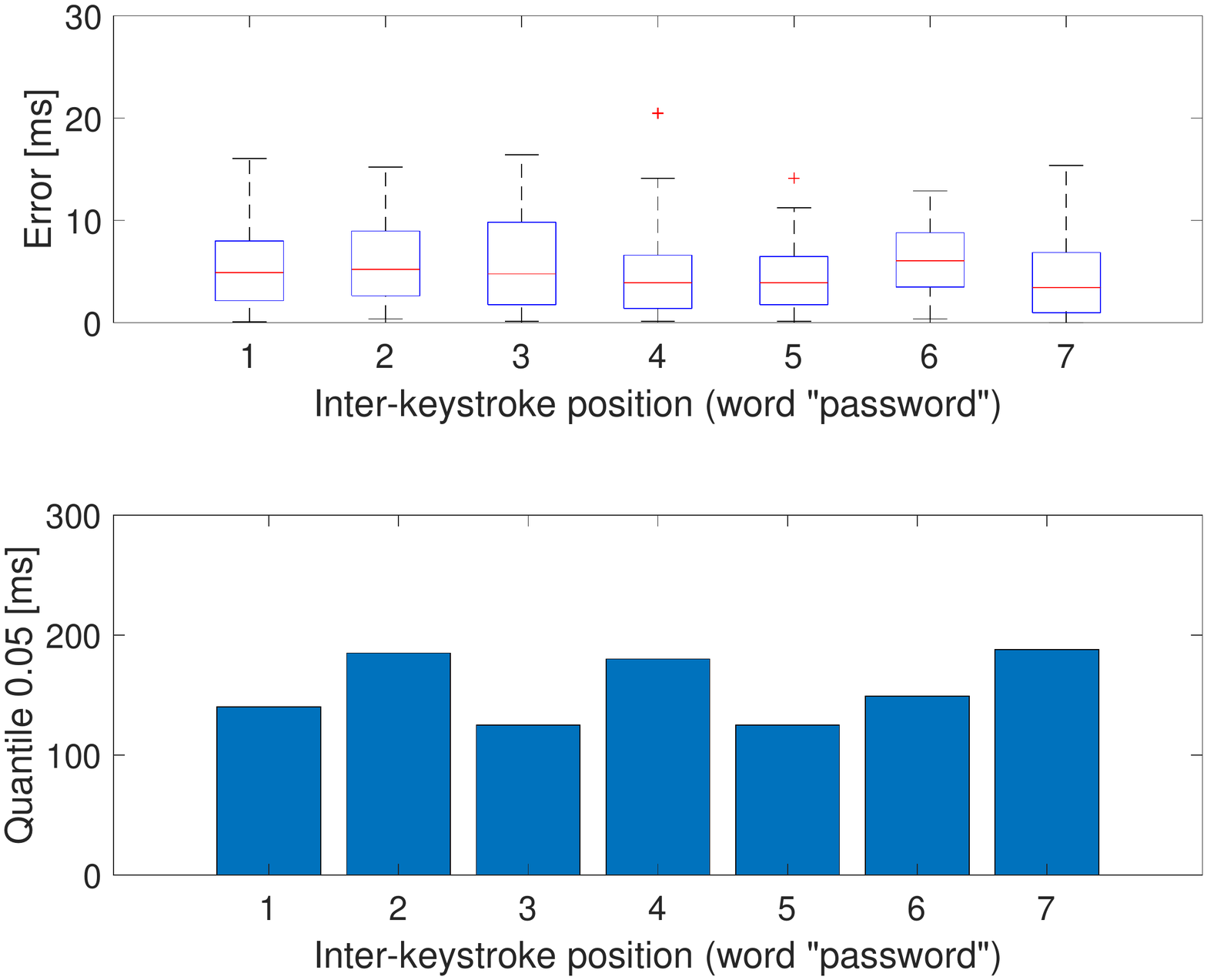}
        \caption{User 3}
    \end{subfigure}
    \caption{Error bound for users $U1$ (a), $U2$ (b), and $U3$ (c), when comparing inter-keystroke timings from a key-logger (bottom figures) with the ones collected by using the \name\ attack and computing the error (top figures).}
    \label{fig:sdr_kl_precision}
\end{figure*}

{\bf Error bounding.} We compare the keystroke timings extracted by the \name\ attack with the timings recorded by a standard keylogger. To this aim, we developed a simple Python script to record the keystroke timings during the previous measurements and, subsequently, we compared such a time sequence with the one collected from the \name\ attack. We performed the previous analysis with three different users, i.e., $U1$, $U2$ and $U3$ as depicted by Fig.~\ref{fig:sdr_kl_precision}. The bottom part of Fig.~\ref{fig:sdr_kl_precision} shows the quantile 0.05 associated with the inter-keystroke timings collected during 50 repetitions of the word ``password'' using the keylogger. In the previous analysis, we did not take into account the carriage-return keystroke, but only the timings between two subsequent keystrokes within the word ``password''. We highlight that we considered only the quantile 0.05 of the keystroke interarrival times, since it represents the worst case, i.e., the keystroke pairs with the 5\% smallest time difference. The top part of Fig.~\ref{fig:sdr_kl_precision} shows the absolute value of the difference (error) between the inter-keystroke timings collected by the \name\ attack and the ones collected by adopting the keylogger (bottom part of Fig.~\ref{fig:sdr_kl_precision}). For each box, the central mark represents the median, while the bottom and top edges of the box indicate the 25th and 75th percentiles, respectively. The whiskers extend to the most extreme data points not considered outliers, and the outliers are plotted individually using the red '+' symbol. We observe that, even in the worst-case scenario, the error is always less than 20ms, compared to an average inter-keystroke timing of 200 ms, computed over the data collected by the key-logger. To sum up, we highlight that the median value of the error is about 5ms (for all the users), being the 2\% of the quantile 0.05 of the inter-keystroke timings collected by the keylogger.

{\bf User independence.} We stress that \emph{Words Identification} and \emph{Keystroke Timings Extraction} are independent of the user, i.e., the processing performed by the \ac{SDR} introduces only a minor delay that does not affect the pattern of the inter-keystroke timings. Therefore, already proposed techniques that affect user's privacy by exploiting inter-keystroke timings, such as the one in~\cite{monaco} and \cite{Song2001}, can be significantly enhanced by moving the adversary far away from the target user. 

\subsection{Keyword detection}
\label{sec:keyword_detection}

Inter-keystroke timings have already been adopted in the literature to infer on patterns and words typed by a target user~\cite{monaco}. Nevertheless, to the best of our knowledge, no one has proposed so far to extract inter-keystroke timings from \ac{RSS} samples. Moreover, our attack significantly improves the chances of the adversary to remain undetected during the guessing procedure. Nevertheless, the combination of the attack peculiarities and the adopted scenarios require a different methodology compared to the ones already proposed in the literature; in particular, we propose a \ac{ML}-based solution that is resilient to both small inter-keystroke timings errors and interference experienced during the eavesdropping phase. 

We considered a \ac{SVM} classifier trained with only one class (one-class \ac{SVM} classifier), i.e., 50 instances of the word ``password''. Our intuition is to discriminate the keyword ``password'' from outliers (other words) by resorting to a likelihood score computed by the \ac{SVM} classifier. The keyword detection phase is performed by the \ac{ML} module of \name, and consists of the following three steps:
\begin{enumerate}
    \item {\bf Training.} We trained a one-class \ac{SVM} model with 50 replicas of the keyword ``password''. We adopted a Gaussian kernel function and we standardized the predictor data, i.e., we centered and scaled each predictor variable by the corresponding weighted column mean and standard deviation; finally, we set the expected proportion of outliers in the training data to 0.05.
    \item {\bf Partition of Inter-Keystroke Timings.} The inter-keystroke timings from the test set are partitioned by using a sliding window with a step size of one keystroke, i.e., two adjacent windows overlap over all the elements but one.
    \item {\bf Score Index Generation.} We test all the partitions using the trained one-class \ac{SVM} classifier obtaining a similarity score (likelihood) for each partition (sliding window).
\end{enumerate}

To either accept or reject a value as the beginning of the keyword, we define a \emph{Decision Threshold}, and the related statistical metrics, i.e., \emph{\ac{TP}}, and \emph{\ac{FP}}.

\begin{proof}[Definition]
    Let $\{s_0, \dots, s_N\}$ be a set of similarity scores. We define {\bf Decision Threshold} ($\Delta$) as the similarity score value such that $min_i(s_i) + \Delta$ represents the minimum value to assume the keyword as included in the sentence.
\end{proof}

\begin{proof}[Definition]
    We define {\bf True Positives (\ac{TP})} the similarity scores that exceed $\Delta$ and, at the same time, feature a position (offset) consistent with the actual position of the keyword in the current sentence.
\end{proof}

\begin{proof}[Definition] 
    We define {\bf False Positives (\ac{FP})} the similarity scores that exceed $\Delta$ and that, at the same time, feature a position (offset) not consistent with the actual position of the keyword. We assume a position as not consistent when its distance from the actual beginning of the keyword is larger than two keystrokes.
\end{proof}
In the next sections, we consider $\Delta=0$ (i.e., we do not consider the effect of $\Delta$), while in Section~\ref{sec:performance}, we study the performance of \name\ for different values of $\Delta$.

It is worth noting that the above-described procedure does not require the attacker to know the time the specific keyword is typed. Indeed, the attacker can first acquire all the keystrokes, and then perform the attack.

\section{Scenario 1: Proximity Attack}
\label{sec:scenario1}

\begin{figure*}
    \begin{subfigure}{0.33\textwidth}
        \includegraphics[width=\linewidth]{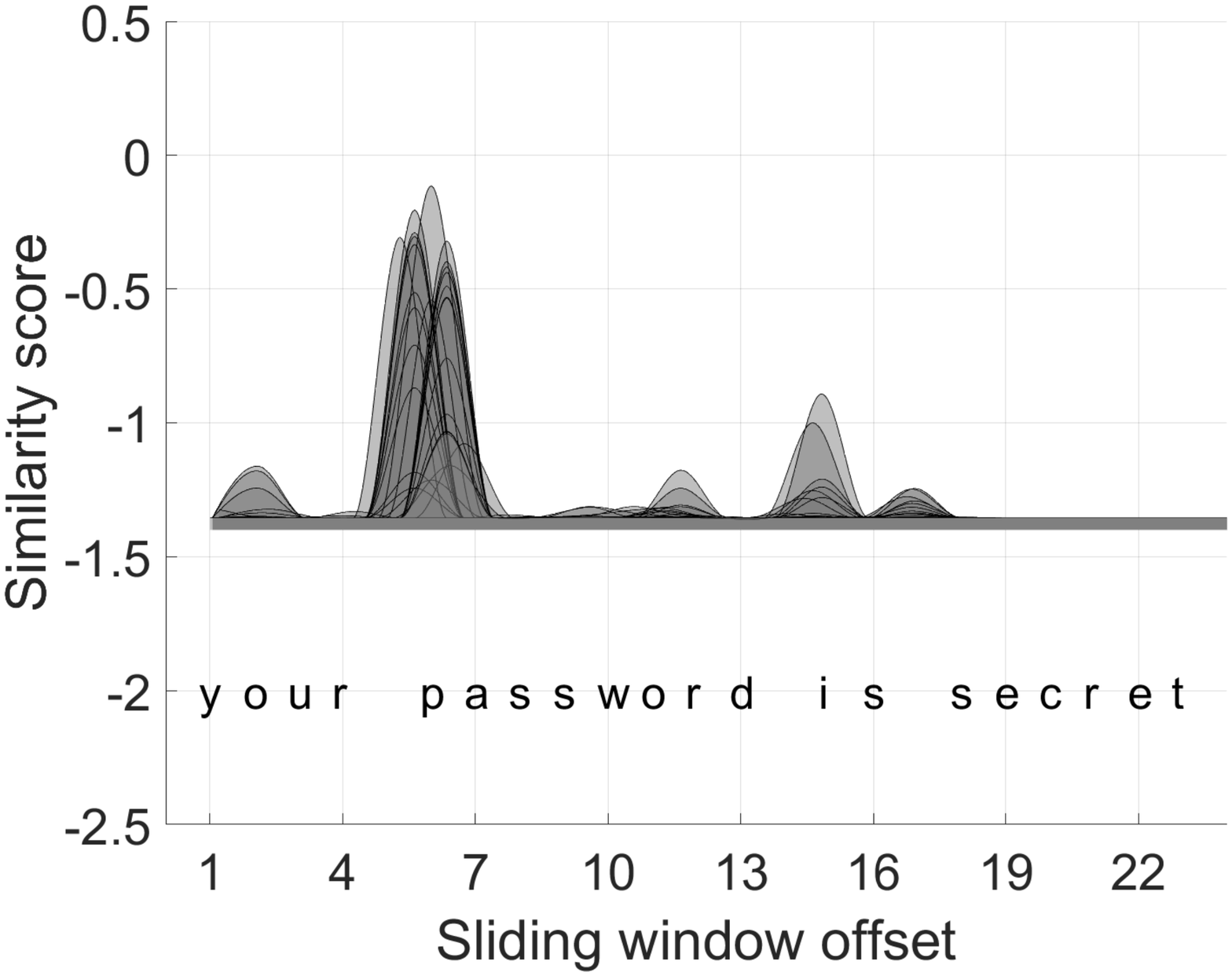}
        \caption{ }
    \end{subfigure}\hfill
    \begin{subfigure}{0.33\textwidth}%
        \includegraphics[width=\linewidth]{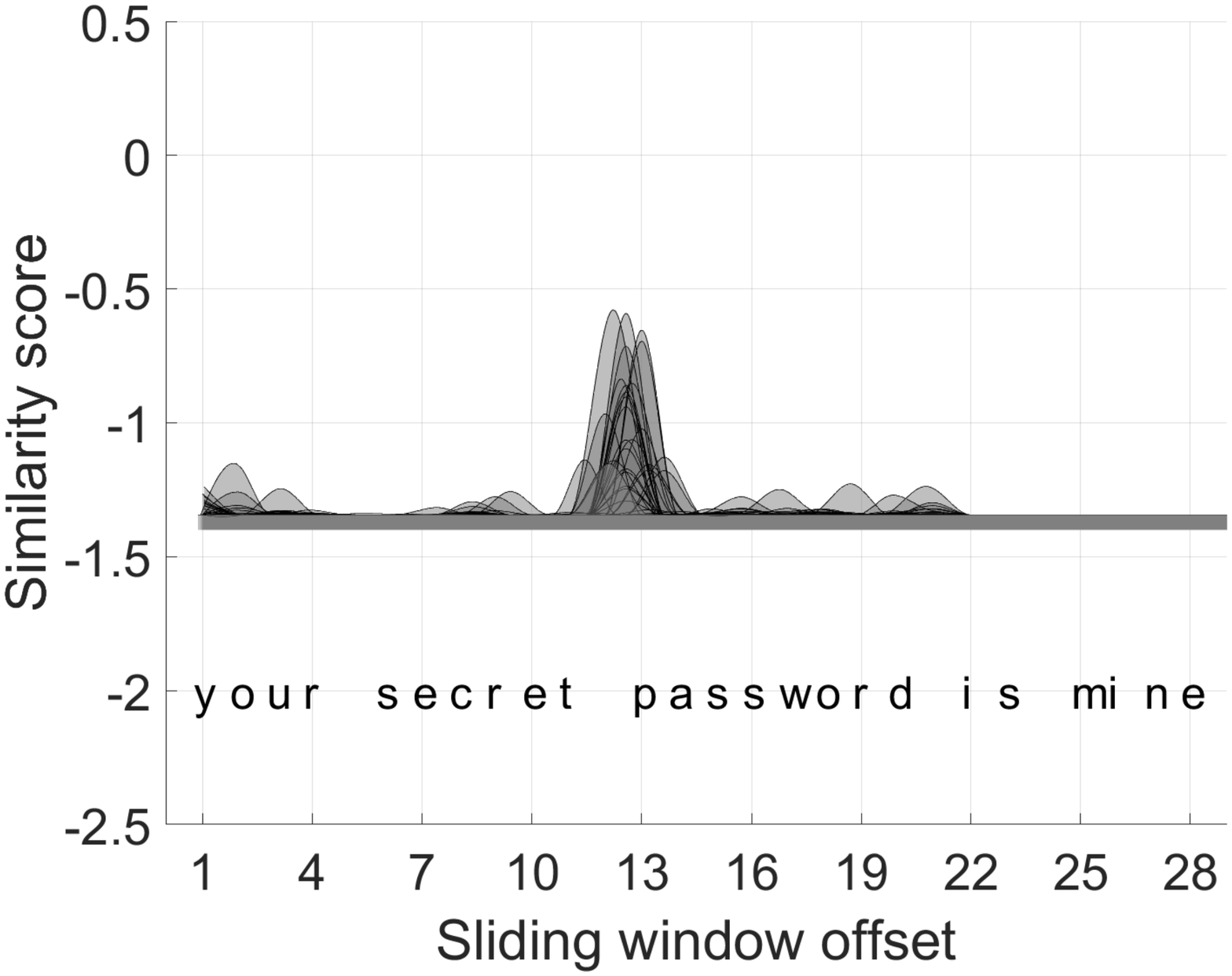}
        \caption{ }
    \end{subfigure}\hfill
    \begin{subfigure}{0.33\textwidth}
        \includegraphics[width=\linewidth]{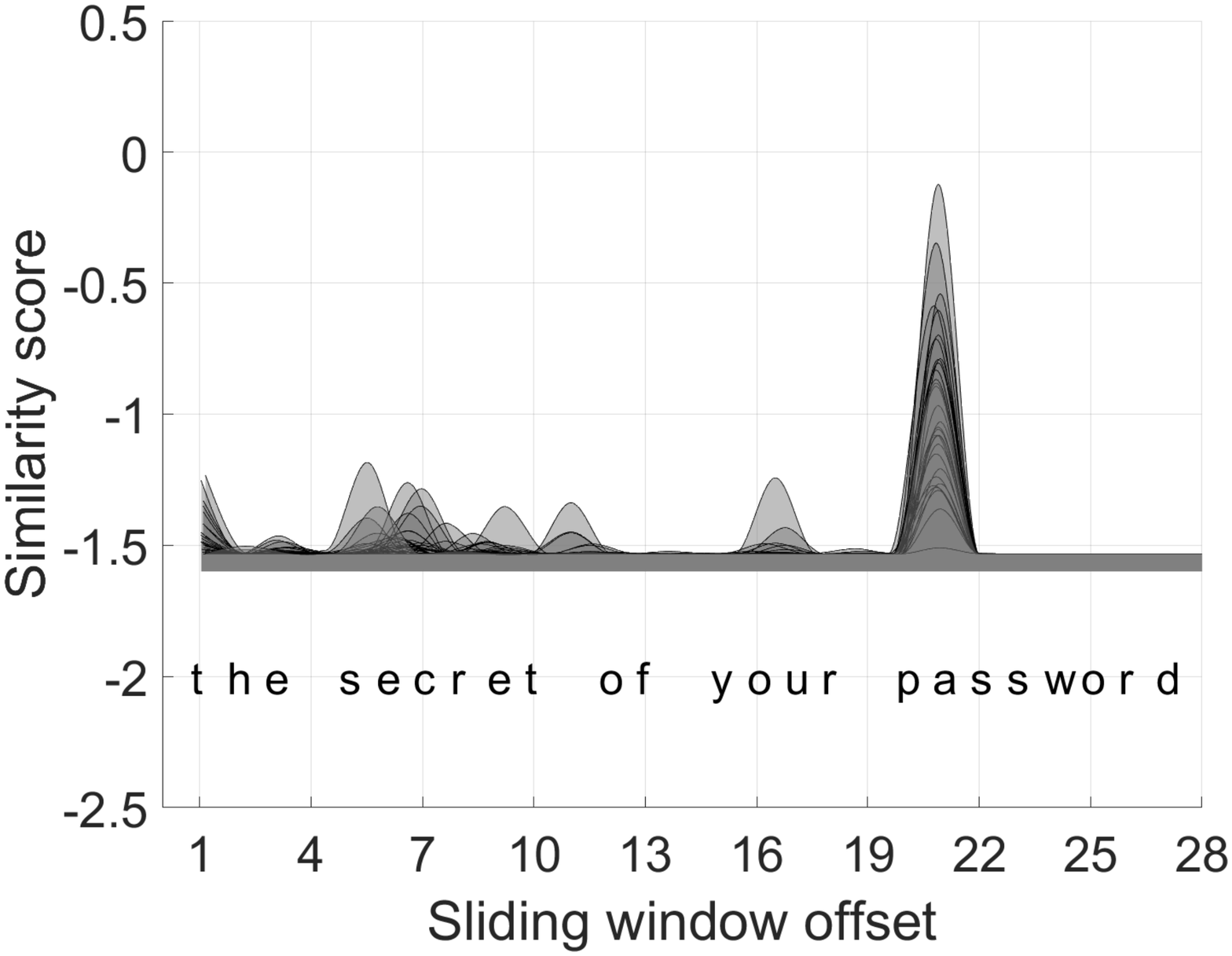}
        \caption{ }
    \end{subfigure}
    \caption{Keyword detection inside a sentence: we tested the \name\ attack against three different sentences (repeated 30 times each) by searching for the keyword ``password''. Similarity scores are generated by the SVM classifier and the peaks represent the likelihood for the beginning of the keyword ``password''.}
    \label{fig:sentence_score}
\end{figure*}

\begin{figure}[htbp]
    \centering
    \includegraphics[width=0.75\columnwidth,height=50mm]{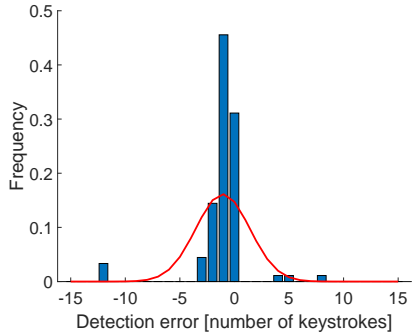}
    \caption{Frequency of the errors associated with the prediction of the keyword position.}
    \label{fig:keyword_position_error}
\end{figure}

We estimate the performance of \name\ in a real-world scenario. We ask user $U1$ to repeat 30 times three different sentences: (i) \emph{your password is secret}; (ii) \emph{the secret of your password}; and, (iii) \emph{your secret password is mine}, being these sentences characterized by the presence of the keyword at different offsets from the beginning of the sentence. We considered Scenario 1 (\emph{Proximity attack}), and therefore we placed the eavesdropping equipment very close to the keyboard-dongle communication link, in a regular office scenario, with people moving around and several sources of interference, i.e., many WiFi networks and Bluetooth devices. Given the proximity between the \ac{SDR} and the keyboard-dongle communication link, we adopted the standard VERT2450 omnidirectional antenna, directly connected to the \ac{SDR}. 
%
%
Figure~\ref{fig:sentence_score} shows the similarity scores provided by the \ac{SVM} classifier as a function of the sliding window offset. The sliding window duration has been calibrated on the number of inter-keystroke timings, i.e., 7, constituting the keyword ``password'' while the sliding step is equal to one keystroke. A peak in the similarity score at a certain offset means that the subsequent samples are likely to match with the samples in the training set, and therefore, the current offset is likely to be the beginning of the keyword. 
We observe that, for all the three sentences, the \ac{SVM} classifier returns higher similarly scores at the offset where the keyword ``password'' begins. Moreover, we observe that \name\ can locate the position of the password, while also experiencing a certain level of uncertainty, i.e., not all the major peaks are located exactly at the position where the keyword begins. 
Indeed, recalling Section~\ref{sec:time_extraction}, interference can either add fake keystrokes or make the existing ones not retrievable. 
Overall, this phenomenon just slightly affects the performance of our attack, and the uncertainty of the keyword position is usually in the range of $\pm 1$ keystroke from the actual position.
By reconsidering the results from Fig.~\ref{fig:sentence_score}, we extracted the maximum score for each sentence and we compared its position with the one corresponding to the actual position associated with the beginning of the keyword ``password''. Figure~\ref{fig:keyword_position_error} shows the number of occurrences as a function of the error in computing the expected position of the keyword. We observe that about 31\% of the detection events do not suffer from any error (27 out of 90). Moreover, we observe that 45\% of the detection events are affected by an error of just one keystroke, while a mere 14\% of the detection events occur 2 keystrokes earlier than the real one. Therefore, \name\ can locate the keyword ``password'' in 90\% of the cases with an error of fewer than 2 keystrokes. A solid red line in Fig.~\ref{fig:keyword_position_error} shows the best fit distribution being a normal distribution with mean -1.06 and standard deviation of 2.47.

\section{Scenario 2: Keyword detection from behind a wall}
\label{sec:scenario2}

In Scenario 2, we perform the attack in an environment characterized by crowded neighboring offices, setting up the eavesdropping equipment in one office and launching the attack from the neighboring office. The target user was aware of our attack and collaborated with us when asked to repeat 30 times the same sentence, i.e., \emph{you can choose a random password}. The antenna has been placed 4.5 meters away from the target user, while a concrete wall of about 20 cm was obstructing the \acl{LOS}. 

\begin{figure}[htbp]
    \centering
    \includegraphics[width=0.75\columnwidth]{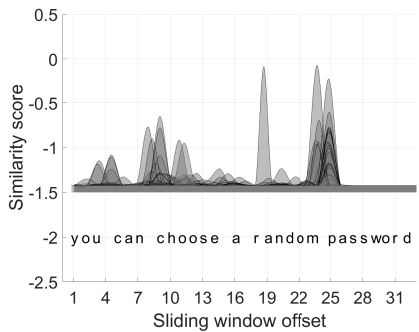}
    \caption{\emph{Behind-the-wall attack} scenario: \name\ is performed against the target keyboard being separated from the eavesdropping antenna by an office wall.} 
    \label{fig:youcanchoosearandompassword}
\end{figure}

We adopted the same measurement setup and analysis as before, and we report the similarity scores in Fig.~\ref{fig:youcanchoosearandompassword} as a function of the sliding window offset. We repeated the previous procedure for a sequence of 30 sentences containing the keyword ``password'' at the $25^{th}$ keystroke.
We observe that the vast majority of the similarity score peaks are concentrated at offsets 24 and 25, i.e., the lag of one keystroke is mainly due to lost samples during the eavesdropping phase. Moreover, we highlight the presence of peaks far away from the expected offset, i.e., one at 19 and a few more in the range from 7 to 13. We consider these peaks as \emph{\ac{FP}s}, i.e., the keyword is not present, but our algorithm still estimated its presence as likely.
Nevertheless, in 19 cases out of 30, the algorithm correctly identifies the position of the keyword, while in 10 cases \name\ provides a (slightly) wrong position for the keyword. 

The number of \ac{FP}s is mainly due to two factors: (i) the wall obstructing the \acl{LOS} affects the \ac{RSS} of the samples transmitted by the keyboard; and, (ii) the office environment is particularly prone to interference. \name\ is particularly sensitive to interference since it exploits \ac{RSS} estimations to generate the inter-keystroke timings. 
We will discuss in detail the strategies to mitigate the number of \ac{FP}s in the next sections.

We notice that, despite all the tested sentences contain the keyword under test \emph{password}, the very low similarity score levels in the positions where the keyword is not present indicate the high robustness of the \name\ attack to \ac{FP}s. In fact, there is no word achieving the same levels of similarity scores as the ones obtained when the keyword is present.

\section{Scenario 3: Remote attack}
\label{sec:scenario3}

In this section, we consider Scenario 3 (\emph{Remote Attack}), where the adversary leverages a directional antenna (Aaronia HyperLOG 60350) to perform the \name\ attack. In this scenario, the target user sits at the ground floor of a two floors villa in Doha, Qatar, in the proximity of a window. We placed the eavesdropping antenna at 1, 5, 10, 15, and 20 meters from the keyboard-dongle communication link. We stress that the link between the directional antenna and the keyboard-dongle is obstructed by only a window, and therefore we consider it as a \ac{LOS} attack.

\begin{figure*}
    \begin{subfigure}{0.2\textwidth}
        \includegraphics[width=\linewidth,height=40mm]{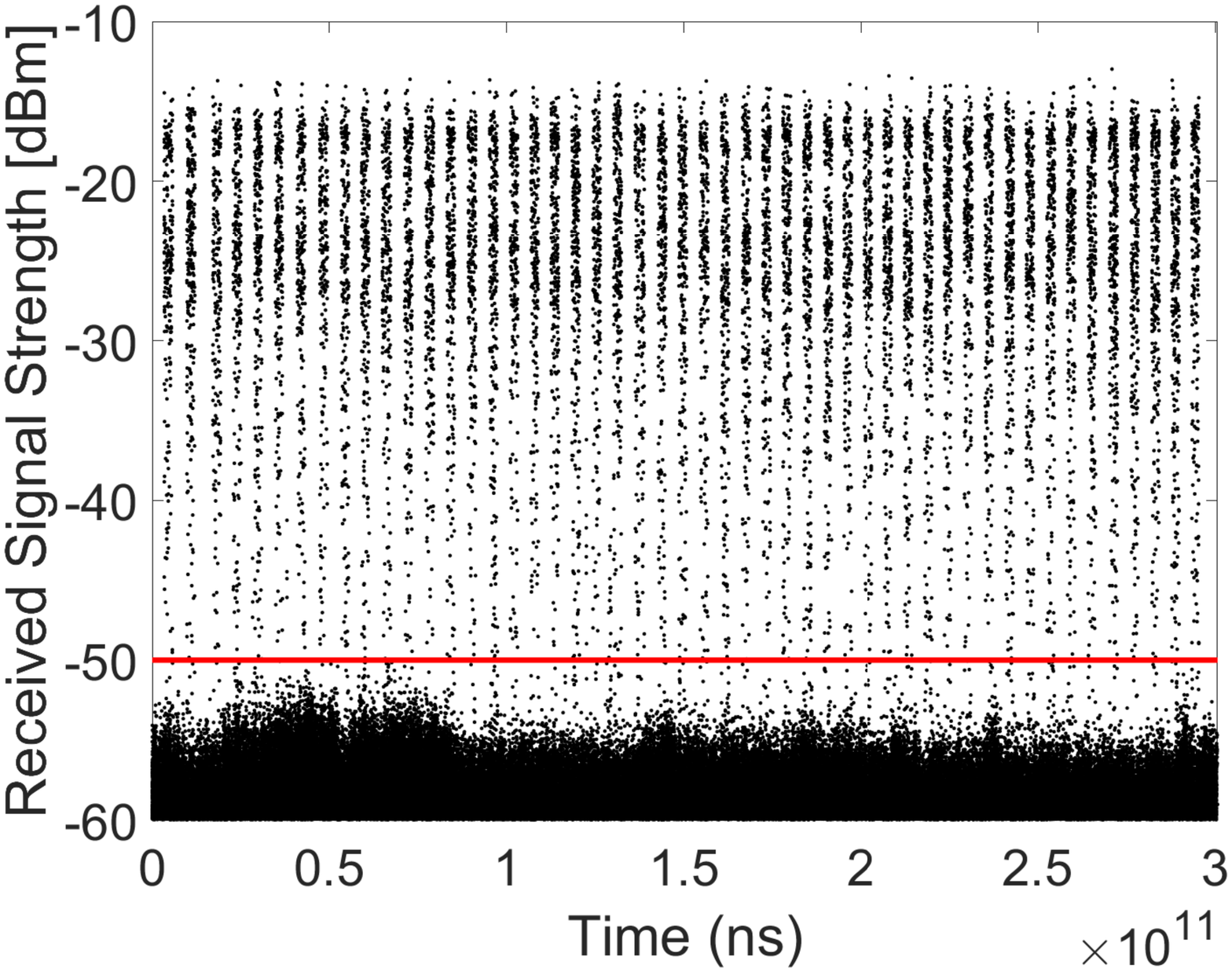}
        \caption{1 meter}
    \end{subfigure}\hfill
    \begin{subfigure}{0.2\textwidth}
        \includegraphics[width=\linewidth,height=40mm]{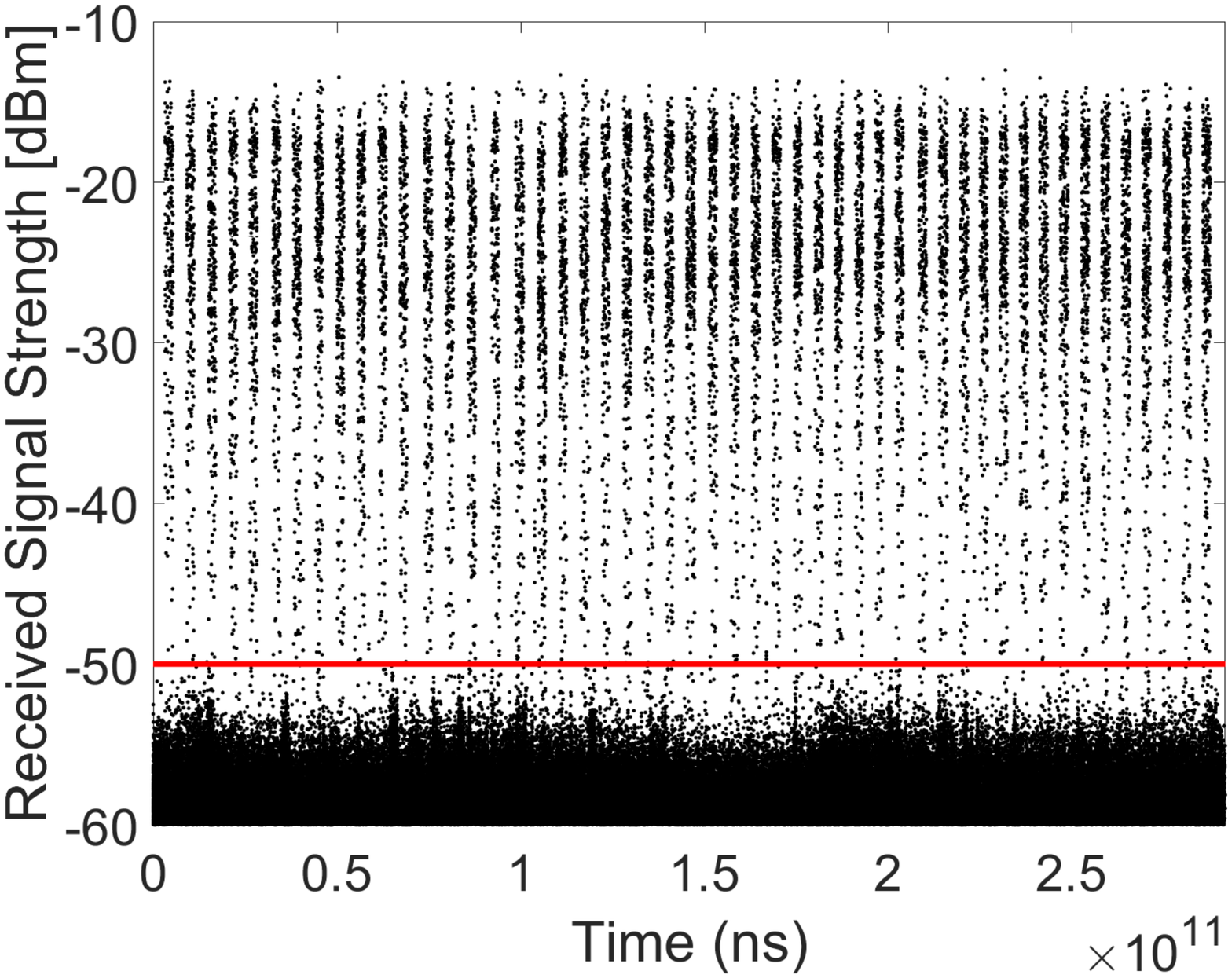}
        \caption{5 meters}
    \end{subfigure}\hfill
    \begin{subfigure}{0.2\textwidth}%
        \includegraphics[width=\linewidth,height=40mm]{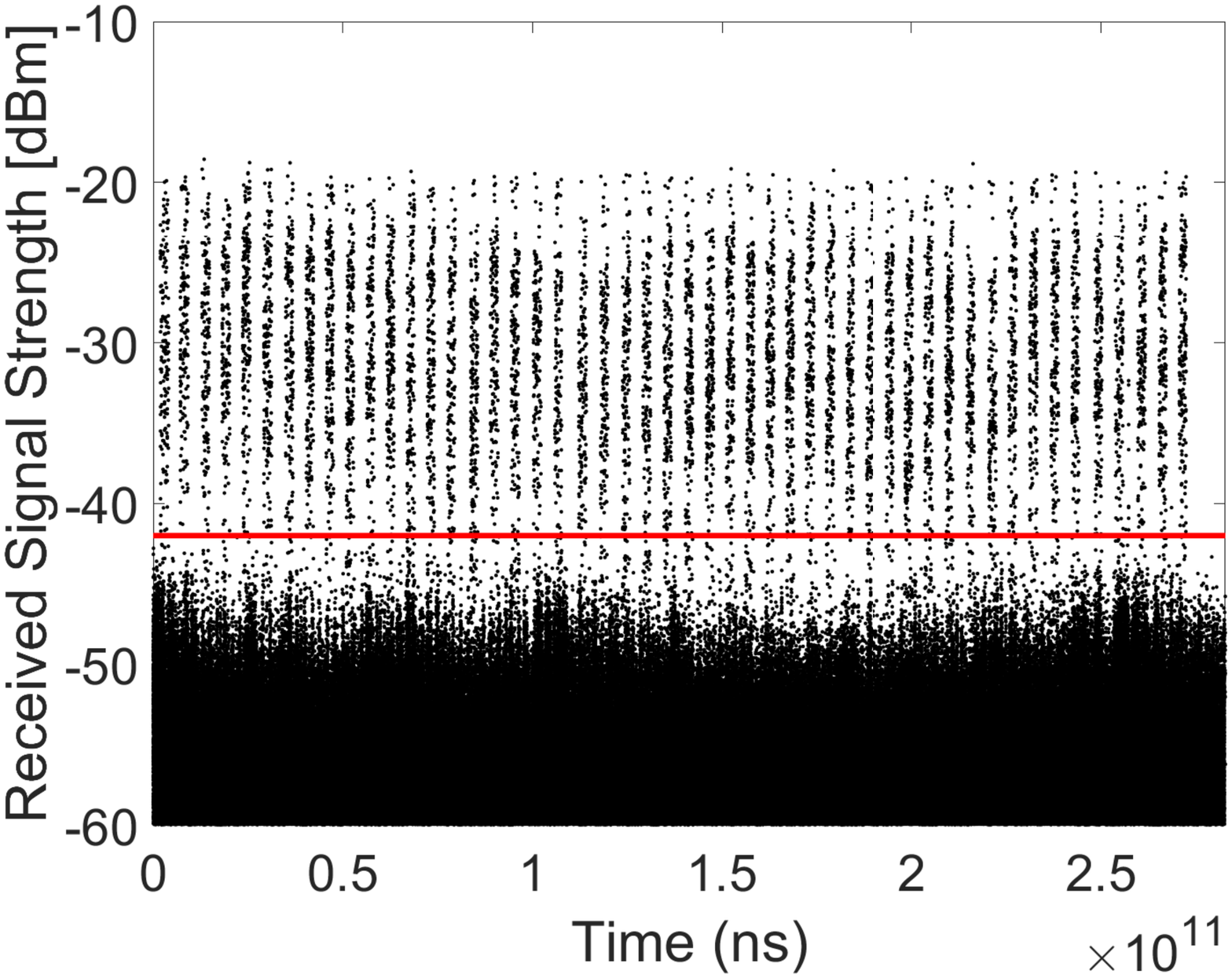}
        \caption{10 meters}
    \end{subfigure}\hfill
    \begin{subfigure}{0.2\textwidth}%
        \includegraphics[width=\linewidth,height=40mm]{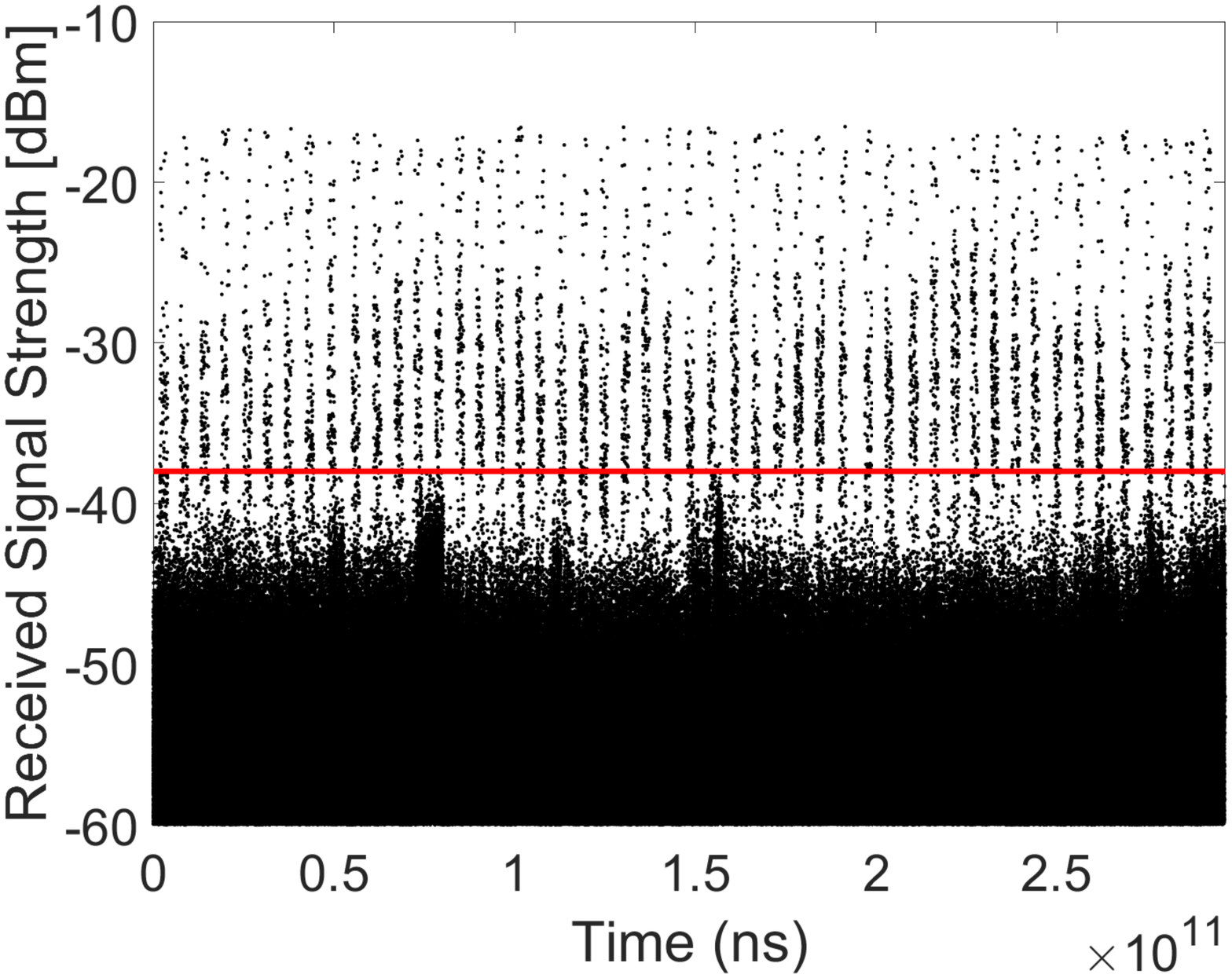}
        \caption{15 meters}
    \end{subfigure}\hfill
    \begin{subfigure}{0.2\textwidth}%
        \includegraphics[width=\linewidth,height=40mm]{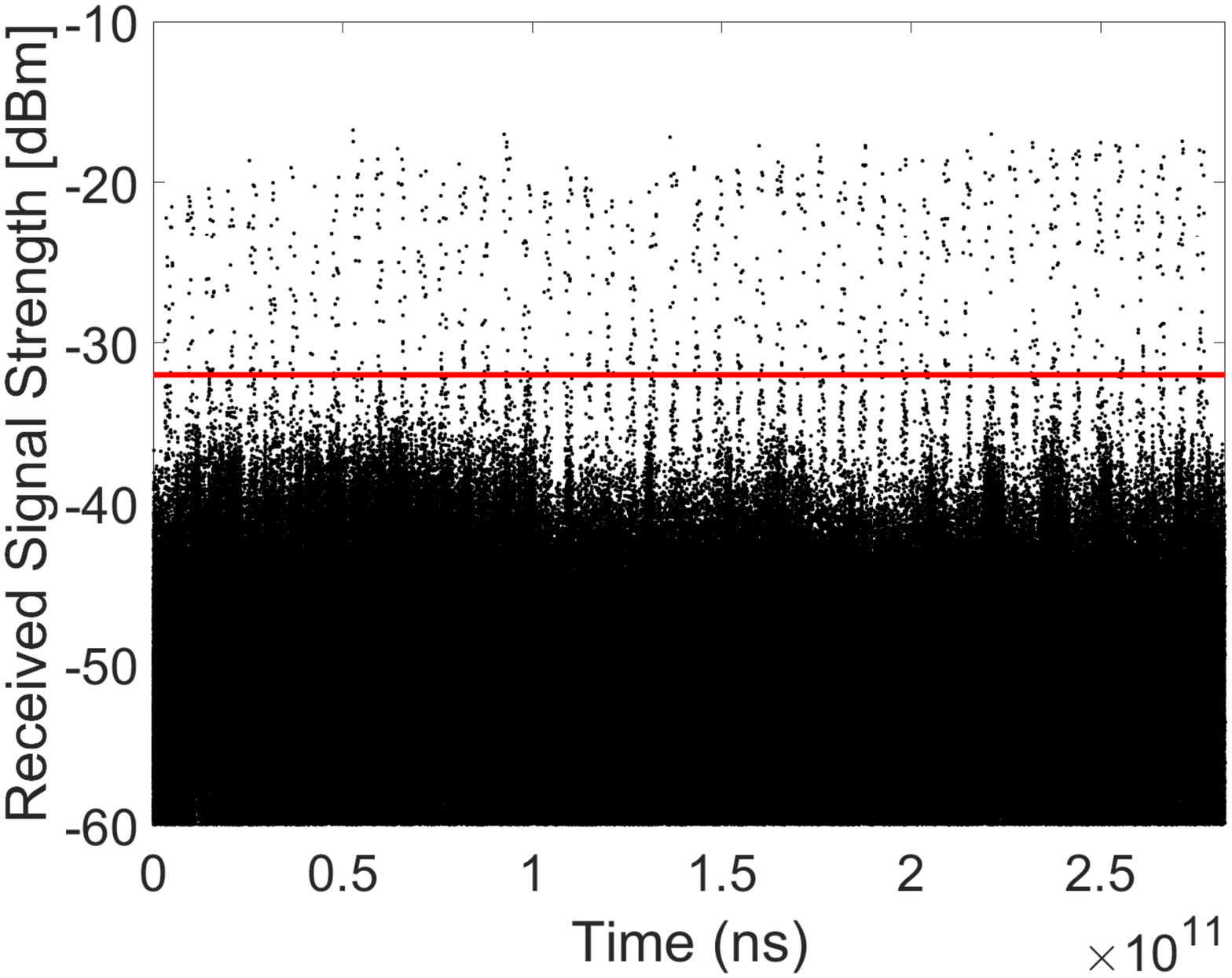}
        \caption{20 meters}
    \end{subfigure}
    \caption{Received Signal Strength (RSS) at  1m, 5m, 10m, 15m, and 20m (from left to right) and related thresholds (red lines) to filter out the noise.\label{fig:distance}}
\end{figure*}

Figure~\ref{fig:distance} shows the \ac{RSS} samples associated with the distances previously considered. Firstly, we observe that the interference significantly increases when the eavesdropping antenna moves away from the target user (black area at the bottom of the figures). This effect can be explained by observing that the main lobe of the directional antenna becomes more and more exposed to transmitting entities that might be located in the neighborhood villas, e.g., WiFi, Bluetooth, and other interfering sources. Moreover, we observe that the peaks associated with the actual \ac{RSS} samples belonging to the keyboard-dongle communication channel are varying between -15dBm and -20dBm at 1m and 20m, respectively. Finally, we calibrated the thresholds (horizontal red lines), by empirically considering the lowest possible values with minimum interference.

\begin{figure}[htbp]
    \centering
    \includegraphics[width=0.8\columnwidth]{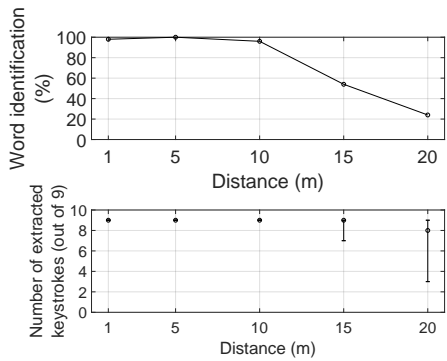}
    \caption{Word extraction ratio (out of 50 repetitions of ``password'') and number of extracted keystrokes (out of 9), with increasing distance. Error bars show quantiles 0.05, 0.5, and 0.95 associated with the number of keystrokes per word extracted from 50 repetitions of ``password''.}
    \label{fig:keywords_distance}
\end{figure}

\begin{figure*}
    \begin{subfigure}{0.33\textwidth}
        \includegraphics[width=\linewidth,height=40mm]{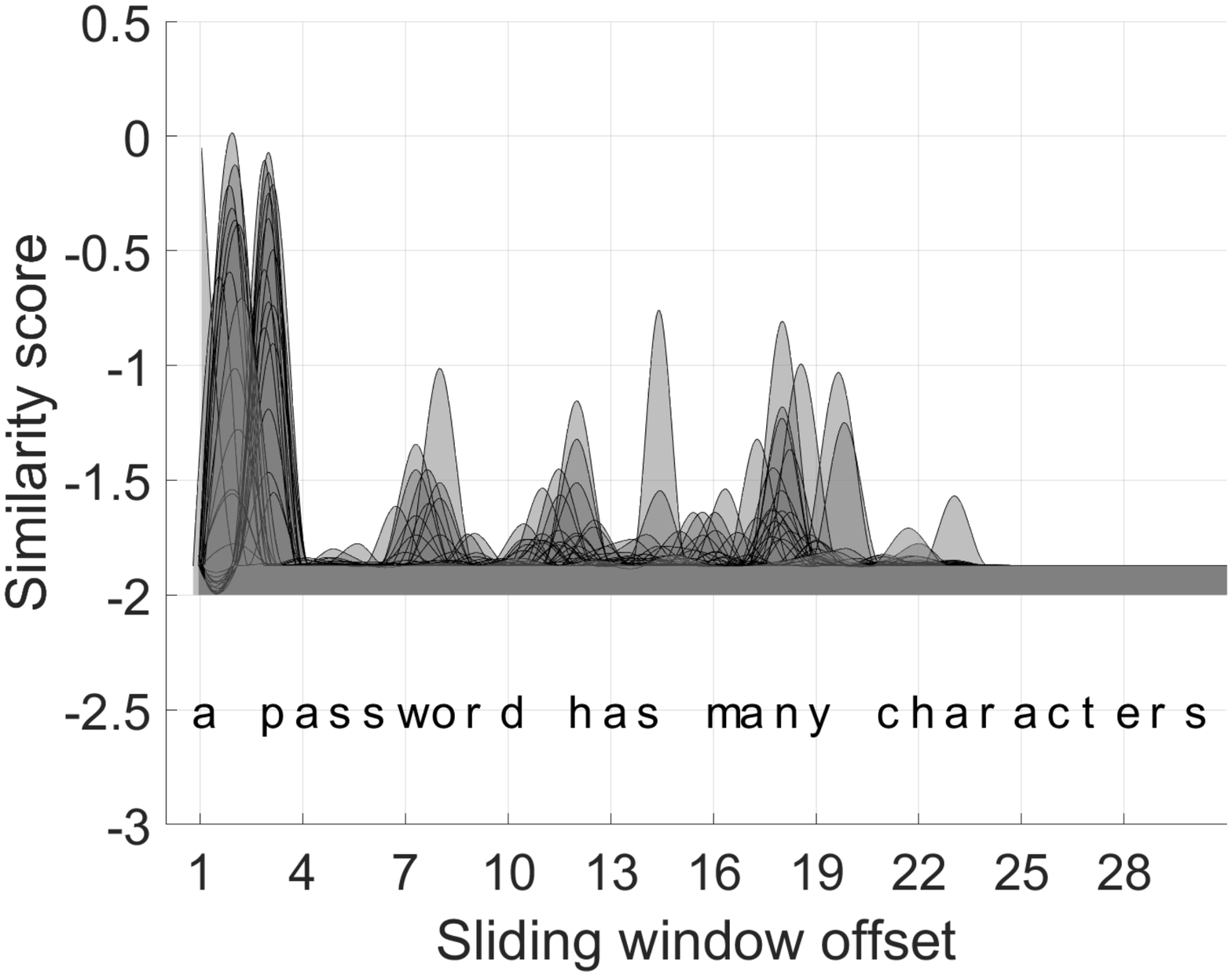}
        \caption{5 meters}
    \end{subfigure}\hfill
    \begin{subfigure}{0.33\textwidth}
        \includegraphics[width=\linewidth,height=40mm]{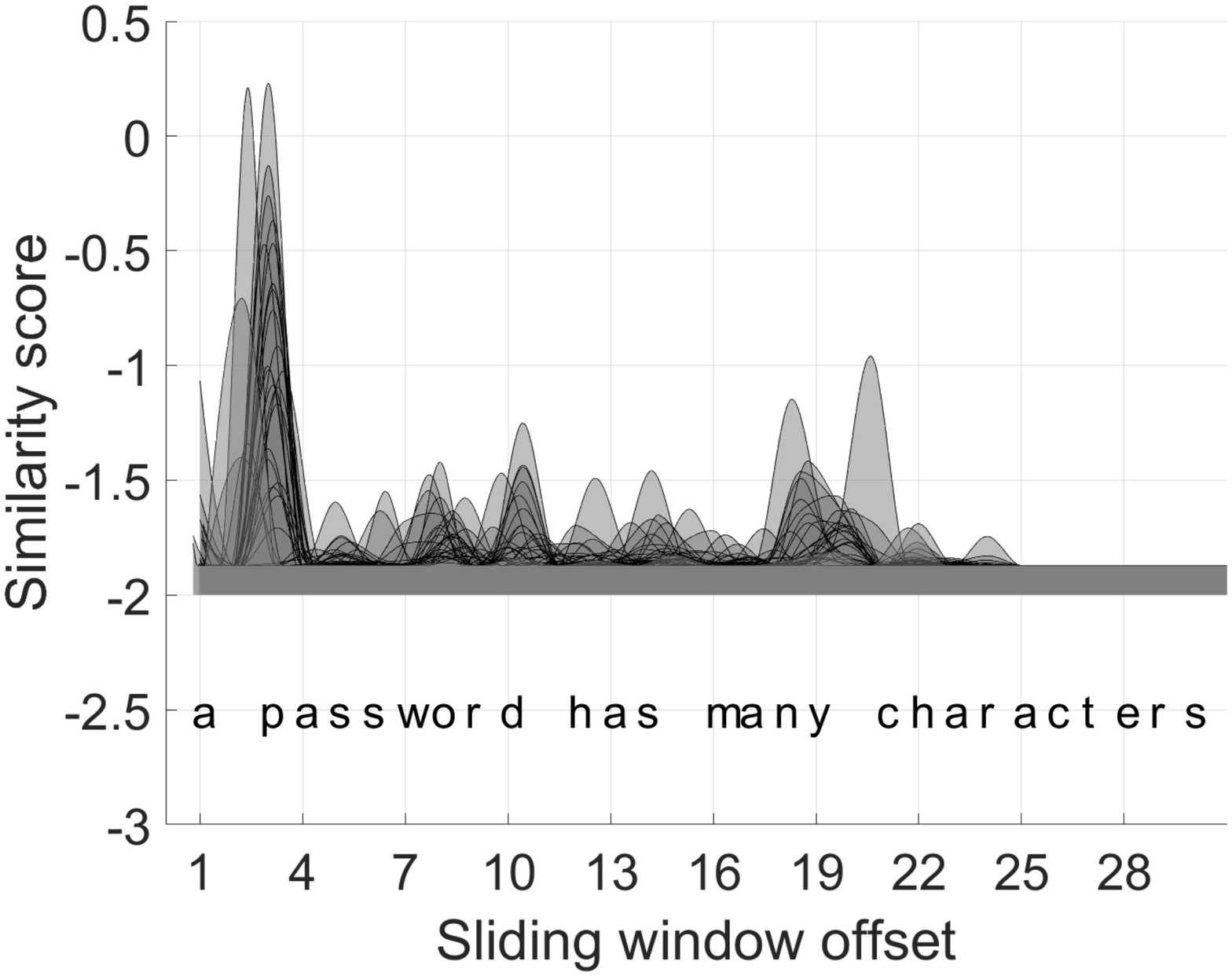}
        \caption{10 meters}
    \end{subfigure}\hfill
    \begin{subfigure}{0.33\textwidth}%
        \includegraphics[width=\linewidth,height=40mm]{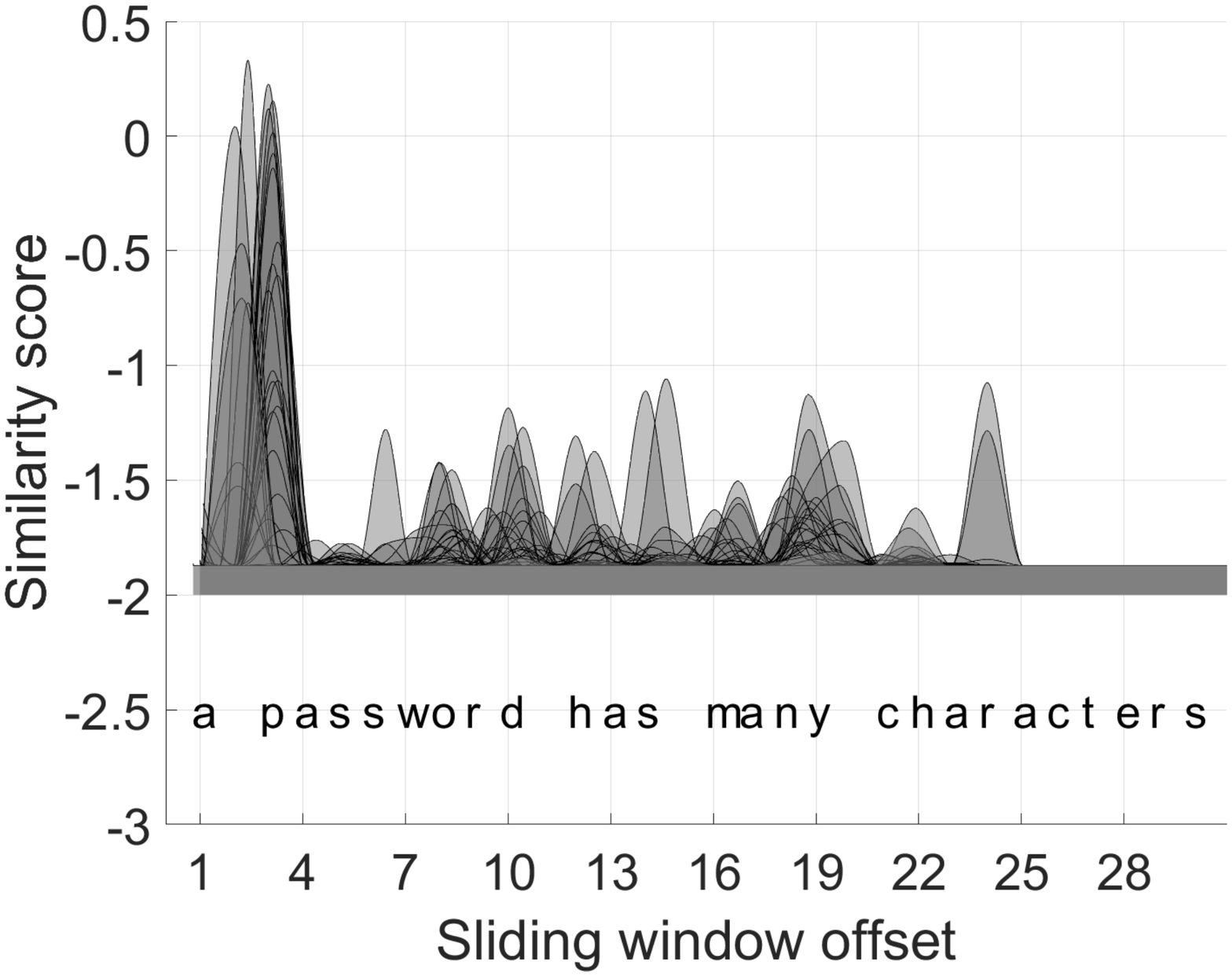}
        \caption{15 meters}
    \end{subfigure}
    \caption{Detecting the keyword ``password'' inside a sentence for Scenario 3 (Remote attack): we tested \name\ against the same sentence (repeated 30 times) at three distances, i.e., 5 meters (a), 10 meters (b), and 15 meters (c). Similarity scores are generated by the SVM classifier and the peaks represent the likelihood for the beginning of the keyword ``password''. \label{fig:remote_sentence}}
\end{figure*}

Figure~\ref{fig:keywords_distance} (top) shows the results of our analysis: \name\ can identify about 100\% of the words up to a distance of 10 meters, while its performance decreases to about 54\%  and 24\% at 15 and 20 meters, respectively. Moreover, Figure~\ref{fig:keywords_distance} (bottom) shows that \name\ can successfully retrieve 9 out of 9 keystrokes (``password'' + carriage return) up to 10 meters, while interference significantly affects performance starting at a distance of 15 meters. Nevertheless, we observe that the number of extracted keystrokes is still high, even at a distance of 20 meters, with a median value of 8 keystrokes identified out of 9. 
\emph{Word Identification} and \emph{Keystroke Timing Extraction} are not enough to detect the presence of the keyword in the keystrokes of the target user. Therefore, in the following, we apply a \ac{ML} technique (\ac{SVM}) to compute the likelihood (similarity score) of the presence (and position) of the keyword ``password'' in a sentence typed by a remote target user, as previously described in Section~\ref{sec:attack_in_brief}. Figure~\ref{fig:remote_sentence} shows the similarity scores generated by the \ac{SVM} algorithm trained with 10 repetitions of the keyword ``password''. Each similarity score is computed by testing a sliding window of 7 inter-keystroke timings with a moving step of 1 keystroke. 
Table~\ref{table:remote_attack} shows the number of \ac{TP}s and \ac{FP}s (out of 30 sentences) as a function of the eavesdropping distance. \name\ can detect the presence and the position of the keyword in the vast majority of the cases ($\ge$73\%), i.e., the peaks of the similarity scores are concentrated at about the same offset ($\pm 1$) of the actual position of the keyword ``password''. Moreover, we observe the presence of a few \ac{FP}s ($\le$23\%), i.e., there are minor peaks distributed at different offsets of the sentence. A thorough analysis of this phenomenon is provided in Section~\ref{sec:performance}.

\begin{table}
    \centering
    \footnotesize
\caption{Remote attack scenario: TP Vs FP \label{table:remote_attack}}
\begin{tabular}{|c|c|c|}
\hline
\multicolumn{1}{|l|}{Distance (m)} & \multicolumn{1}{c|}{TP} & \multicolumn{1}{c|}{FP} \\ \hline
5                                  & 24/30                      & 5/30                       \\ \hline
10                                 & 22/30                      & 7/30                       \\ \hline
15                                 & 24/30                      & 5/30                       \\ \hline
\end{tabular}
\end{table}

\section{\name\ performance}
\label{sec:performance}

\begin{figure}[htbp]
    \centering
    \includegraphics[width=0.8\columnwidth,height=50mm]{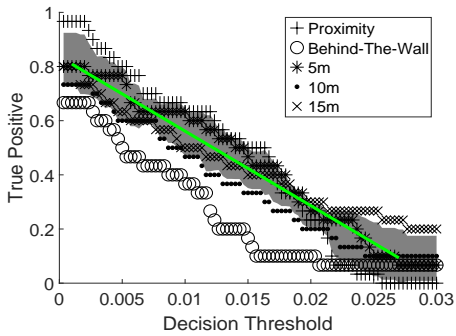}
    \caption{\ac{TP}s increasing the Decision Threshold.}
    \label{fig:TPThreshold}
\end{figure}

In this section, we provide an estimation of the \name\ attack performance, considering all the three discussed scenarios altogether. As previously detailed, we trained a statistical learning algorithm (\ac{SVM}) with a sequence of 10 repetitions of the keyword ``password'', and we tested such a model on adjacent subsets (sliding windows) of several sentences. 
For each test, the \ac{SVM} algorithm provides a similarity score (i.e., likelihood) that such a subset of characters matches the keyword we are looking for. Therefore, each sentence becomes a vector of similarity scores. 
In previous sections, we considered a decision threshold $\Delta=0$, while in the following, we study how $\Delta$ affects the performance of \name---we will vary $\Delta$ in the range  $[0,\ldots,0.03]$. Figure~\ref{fig:TPThreshold} shows the \ac{TP}s estimations as a function of $\Delta$. We consider all the major measurements we already discussed in this paper: (i) Proximity attack (sentence ``your password is secret''); (ii) Behind-The-Wall attack  (sentence ``you can choose a random password''); and, (iii) Remote attack at distances of 5, 10, and 15 meters (sentence ``a password has many characters''), respectively. We didn't consider 1m and 20m: the former has performance similar to Scenario 1 (\emph{Proximity Attack}), while the latter one is affected by too much interference. Firstly, we observe that the Proximity and Remote attacks are characterized by similar trends that can be modeled with a straight line with slope -27 and Y-intercept equal to 0.83648 (solid green line).
We also observe that the worst performance are from the Behind-The-Wall scenario; as previously discussed, such scenario is the only one deprived of the \ac{LOS}, while at the same time suffering from interference generated by neighboring devices. Lastly, we highlight that \name\ can detect the presence of a keyword in the inter-keystrokes timings samples of a target user with a frequency of about 80\%, independently of the considered scenario.

\begin{figure}[htbp]
    \centering
    \includegraphics[width=0.8\columnwidth,height=50mm]{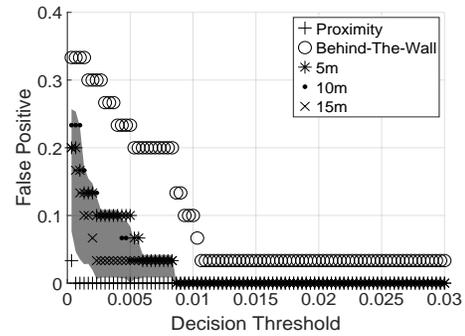}
    \caption{\ac{FP}s increasing the Decision Threshold.}
    \label{fig:FPThreshold}
\end{figure}

Figure~\ref{fig:FPThreshold} shows the \ac{FP}s estimations as a function of the Decision Threshold ($\Delta$). As for the previous case, we consider the: (i) Proximity attack (sentence: ``your password is secret''); (ii) Behind-The-Wall attack (sentence: ``you can choose a random password''); and, (iii) Remote attack at distances of 5, 10, and 15 meters (sentence: ``a password has many characters''), respectively. 
Figure~\ref{fig:FPThreshold} confirms that the Behind-The-Wall scenario is the least performing: for all the thresholds, the \ac{FP}s in this scenario are significantly higher than the ones in the other scenarios (although being always less than 35\%). Conversely, the other scenarios show better performance, being characterized by some \ac{FP}s---always less than 25\%.
The Decision Threshold ($\Delta$) value should be chosen to maximize the number of \ac{TP}s, while at the same time reducing the number of \ac{FP}s. Nevertheless, given the results of Fig.~\ref{fig:TPThreshold} and Fig.~\ref{fig:FPThreshold}, we observe that $\Delta$ should be chosen as small as possible ($<5 \cdot 10^{-3}$) to experience high values of \ac{TP}s, and therefore, low values of missed detection (\ac{FN}s). Conversely, the number of \ac{FP}s can be estimated as 10\% (on average) when $\Delta < 5 \cdot 10^{-3}$; we deem this values as  an acceptable one, since we can assume one or more additional layers of post-processing to reduce the number of false alarms, by exploiting advanced \ac{ML} techniques---though left for future work, we highlight the issue in next section. The source data adopted by this work have been released as open-source at the link \cite{dataset}, to allow practitioners, industries, and academia to verify our claims and use them as a basis for further development.

\section{Discussion}
\label{sec:discussion}

In the following, we discuss the importance of the training set size, some limitations of \name\ and, finally, a few potential countermeasures to mitigate its impact.

{\bf Training set size.} The effectiveness of \name\ strongly relies on the training set previously collected by the adversary. On the one hand, large training sets might be difficult to collect in a reasonable amount of time, and therefore, the attack feasibility is strictly related to the number of required repetitions of the keyword to achieve good detection performance. On the other hand, small training sets can be easily collected by simple social engineering techniques, for instance triggering a response from the user (i.e., having his typing) via e-mail or social networks, to cite a few. We studied the performance of the \name\ attack with different training set sizes, from 5 to 50 repetitions of the keyword ``password''. We considered the 30 repetitions of the sentence ``your password is secret'' from Scenario 1 (\emph{Proximity Attack}) as our test set, and we run the \name\ attack as described in the previous sections. The optimal size of the training set is 10, guaranteeing the maximum number of \ac{TP}s (29) and minimizing the number of \ac{FP}s (1), as depicted in Table~\ref{table:train_set_size}.
\begin{table}[htbp]
    \centering
\caption{\ac{TP} and \ac{FP} as a function of the training set size.}
\begin{tabular}{lcccccc}
 & \multicolumn{6}{c}{\textbf{Train Set Size}} \\ 
\multicolumn{1}{l|}{} & \multicolumn{1}{l|}{\textbf{5}} & \multicolumn{1}{l|}{\textbf{10}} & \multicolumn{1}{l|}{\textbf{20}} & \multicolumn{1}{l|}{\textbf{30}} & \multicolumn{1}{l|}{\textbf{40}} & \multicolumn{1}{l|}{\textbf{50}} \\ \hline
\multicolumn{1}{l|}{\textbf{TP}} & \multicolumn{1}{c|}{25} & \multicolumn{1}{c|}{29} & \multicolumn{1}{c|}{28} & \multicolumn{1}{c|}{26} & \multicolumn{1}{c|}{26} & \multicolumn{1}{c|}{24} \\ \hline
\multicolumn{1}{l|}{\textbf{FP}} & \multicolumn{1}{c|}{5} & \multicolumn{1}{c|}{1} & \multicolumn{1}{c|}{2} & \multicolumn{1}{c|}{4} & \multicolumn{1}{c|}{4} & \multicolumn{1}{c|}{6} \\ \hline
\end{tabular}
\label{table:train_set_size}
\end{table}
The training set size leading to the best results depends on the keyword and the user typing pace. Thus, a preliminary phase is required to estimate the optimal training set size for each keyword-user combination.\\
{\bf Keyboard communication protocol.} 
The vast majority of keyboards adopt \emph{proprietary protocols}, like the ones used throughout this paper. These protocols usually select a frequency and keep it for a long-term period (up to the switch-off or battery replacement). This behavior is particularly prone to the \name\ attack, since the attacker can monitor the ISM band, 
in the range 2.4-2.5 GHz, identify the frequency adopted by the target user, and select that target frequency for collecting the RSS samples.

\begin{figure}[htbp]
    \centering
    \includegraphics[width=0.8\columnwidth,height=50mm]{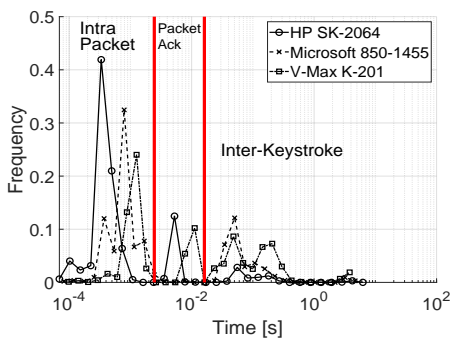}
    \caption{Keyboards comparison.
    }
    \label{fig:keyboards}
\end{figure}

Figure~\ref{fig:keyboards} shows the inter-sample timings for the three different keyboards discussed in Section~\ref{sec:scenario}. We distinguish three categories: (i) Intra-packet samples; (ii) Packet-Ack delay; and, (iii) Inter-keystroke timings. Intra-packet samples are the RSS estimations belonging to the same packet, being either the message from the keyboard to the dongle or the acknowledgment from the dongle to the keyboard. The second category (Packet-Ack delay) is the time between the packet and the ack: the keyboard Microsoft 850-1455 seems to have a very small delay compared to both HP and V-Max. We consider 16ms as the upper bound for the previous category. Finally, Inter-Keystroke timings represent the time between two consecutive keystrokes. \name\ is effective if and only if the user's typing speed is lower than the Packet-Ack delay. When the user's typing speed becomes comparable to the Packet-Ack delay, the current version of \name\ is not able to distinguish between the Ack of a packet and the packet associated with the subsequent keystroke. We recall that we empirically chose 23ms (Section~\ref{sec:keystroke_timing_extraction}) to uniquely identify the Packet-Ack pattern for the keyboard HP SK-2064.

Keyboards adopting Bluetooth, and therefore frequency hopping, require a larger spectrum observation to capture the \ac{RSS} samples of the pseudo-randomly chosen frequencies, thus increasing the cost of the equipment used to launch our attack. Moreover, some other keyboard producers (a negligible fraction of, though) adopt the \ac{DSSS} modulation, which spreads the information over a wide-band channel, significantly reducing the transmission peak power and making the communication almost indistinguishable from the noise floor. 

{\bf External interference.} The main drawback to \name\ is interference. As previously discussed, other devices sharing the same frequencies of the keyboard-dongle communication link might significantly affect the performance of the attack. We studied the effect of interference, by considering different parameters (i.e., \ac{RSS} thresholds), equipment (i.e., directional and omnidirectional antenna), and scenarios (i.e., Proximity, Behind-The-Wall, and Outdoor). We proved that interference can be mitigated and \name\ can guarantee the detection of a keyword with high chances (more than 70\% in the harshest conditions), independently of the configuration.

{\bf Countermeasures.} To mitigate  \name, the following strategies could be implemented: (i) increasing the number of transmissions by either beaconing or friendly jamming; (ii) randomly delaying the keyboard transmissions; or, (iii) adopting \ac{DSSS} instead of fixed or pseudo-random frequency hopping techniques. The first two strategies might be impractical for wireless keyboards, since they require more energy, with the second one also possibly affecting the user experience. Wireless keyboards are mainly event-triggered devices and the trade-off between energy, usability, and privacy has already been widely investigated~\cite{leu2018send}. Finally, although \ac{DSSS} might appear an effective strategy, it is more energy-consuming than frequency hopping~\cite{Lopelli2011}, leading to consider 
a trade-off between privacy objectives and energy budget. 

\section{Conclusion}
\label{sec:conclusion}

In this paper, we have introduced \name, a novel, inexpensive, viable, efficient, and effective attack targeting commercial wireless keyboards.  \name~allows  to detect the presence of a pre-defined keyword in a stream of user-generated keystrokes, by just analyzing the wireless traffic generated by the keyboard.

We studied the effectiveness of \name\ in three different scenarios, including proximity to the target user, \ac{LOS} with distances spanning between 1 and 15 meters, and non-LOS scenarios (eavesdropping from behind a wall in a crowded office environment). All the scenarios are characterized by remarkable performance even in the presence of noise (from more than $70\%$ in the harshest conditions to 90\%+ in normal operating conditions), confirming both the viability and effectiveness of the attack. We also highlighted some limitations, as well as future interesting research directions.

%

\section*{Acknowledgements}
This publication was partially supported by awards NPRP-S-11-0109-180242, NPRP12S-0125-190013, and NPRP X-063-1-014 from the QNRF-Qatar National Research Fund, a member of The Qatar Foundation. The information and views set out in this publication are those of the authors and do not necessarily reflect the official opinion of the QNRF.

\bibliographystyle{IEEEtran}
\balance
\bibliography{biblio}
\balance

\end{document}